\documentclass[11pt,a4paper]{article}
\usepackage[utf8x]{inputenc}
\usepackage[T1]{fontenc}
\usepackage{mathptmx} 

\usepackage[margin=25truemm]{geometry}
\usepackage[pdftex]{graphicx} 
\usepackage{calc} 
\usepackage{authblk}
\usepackage{bm}
\usepackage{enumitem} 
\usepackage{braket,mathtools,amsmath,amssymb,subcaption}
\usepackage{authblk}
\usepackage[all]{nowidow} 
\usepackage{url}

\usepackage{lipsum} 

\newcommand{\fl}[1]{^{({#1})}}
\newcommand{\fii}[1]{^{I({#1})}}
\newcommand{\tl}[1]{\Tilde{#1}}
\newcommand{\tlm}[1]{\Tilde{\lambda}_{#1}}

\usepackage{framed} 
\usepackage[framed]{ntheorem}
\newframedtheorem{frm-thm}{Theorem}

\def\tht{'t Hooft}

\def\ta{\Tilde{a}}
\def\tb{\Tilde{b}}
\def\tA{\Tilde{A}}
\def\tc{\Tilde{c}}
\def\hx{\hat{x}}
\def\hy{\hat{y}}
\def\hz{\hat{z}}

\def\br{\mathbf{r}}
\def\bi{\mathbf{l}_I}
\def\bx{\mathbf{l}_x}
\def\by{\mathbf{l}_y}
\def\bz{\mathbf{l}_z}

\def\fr{\frac{1}{2}}

\def\wde{\wedge}
\usepackage{hyperref}
\hypersetup{colorlinks,linkcolor=blue,citecolor=magenta}
\begin{document} 

\title{Anomaly inflow for dipole symmetry and higher form foliated field theories}

\author{Hiromi Ebisu$^1$}
\author{Masazumi Honda$^2$}
\author{Taiichi Nakanishi$^{2,1}$}
\affil{$^1$Yukawa Institute for Theoretical Physics, Kyoto University, Kyoto 606-8502, Japan}
\affil{$^2$Interdisciplinary Theoretical and Mathematical Sciences Program (iTHEMS), RIKEN, Wako 351-0198, Japan}
                            
\maketitle
\thispagestyle{empty}

\begin{abstract}
In accordance with recent progress of fracton topological phases, unusual topological phases of matter hosting fractionalized quasiparticle excitations with mobility constraints, new type of symmetry is studied -- \textit{multipole symmetry}, associated with conservation of multipoles. 
Based on algebraic relation between dipole and global charges, 
we introduce a series of $(d+1)$-dimensional BF theories with $p$-form gauge fields, which admit dipole of spatially extended excitations,  
and study their physical properties.
We elucidate that gauge invariant loops have unusual form, containing linear function of the spatial coordinate, which leads to the position dependent braiding statistics and unusual ground state degeneracy dependence on the system size. 
We also show that the theories exhibit a mixed \tht~anomaly between $p$-form and $(d-p)$-form dipole symmetries, which is canceled by an invertible theory defined in one dimensional higher via anomaly inflow mechanism.

\end{abstract}

\newpage
\pagenumbering{arabic}

\tableofcontents
\section{Introduction}
Symmetry plays a prominent role in physics, and have been a guiding principle for many researches in various contexts. Here we list some examples: prediction of the form of energy spectrum, classifying phases of matter based on symmetry breaking, understanding of physics in the vicinity of fixed points of critical phenomena, and many more. 
In recent years, a plethora of efforts have been devoted to update the concept of symmetries. 
One such example is \textit{generalized symmetry}, a.k.a., \textit{higher form symmetry}~\cite{doi:10.1073/pnas.0803726105,nussinov2009symmetry,gaiotto2015generalized}, which is associated with symmetry of extended objects. Fruitful interdisciplinary advances have been made for decades by studying this symmetry, allowing us to make better understanding of problems not only in the context of high energy physics but also in view of condensed matter physics, such as classifying topological phases of matter.\footnote{%
See, e.g., recent pedagogical reviews~\cite{mcgreevy2023generalized,bhardwaj2024lectures} and references therein.}\par

Newly proposed fracton topological phases~\cite{chamon,Haah2011,Vijay} have motivated one to explore other types of symmetries -- \textit{multipole symmetries}. 
To see how, let us first briefly recall the property of the fracton topological phases.
Distinctive feature of these phases is that mobility constraints are imposed on fractionalized quasiparticle excitations, giving rise to subextensive ground state degeneracy (GSD) on a torus geometry~\cite{PhysRevX.8.031051,shirley2019fractional}. Due to this property, one faces an issue with the UV/IR mixing, implying incapability to resort to preexisting theoretical frameworks, such as topological field theories. To handle this problem, new types of symmetries have been introduced.
One of such symmetries is multipole symmetry~\cite{griffin2015scalar,Pretko:2018jbi,PhysRevX.9.031035,PhysRevB.98.035111,PhysRevB.106.045112,Jain:2021ibh,hirono2022symmetry}, associated with conservation of multipole moments, such as dipole, quadrupole, and octopole, e.t.c. 
Previously, it was argued that the multipole symmetries give rise to mobility constraints on excitations, which play a crucial role in understanding of fracton topological phases~(see, for instance, \cite{PhysRevB.98.035111,PhysRevB.97.235112,ebisu2209anisotropic,delfino2023anyon,han2024dipolar}). 
Furthermore, based on the algebraic relations between global and dipole charges, a series of topological models have been constructed~\cite{2023foliated,2024multipole}. 
Despite several attempts, full understanding of these symmetries remains elusive. 
\par
In this paper, we discuss a class of field theories with dipole symmetry, which is the simplest example of the multipole symmetries, and their anomalies. 
Furthermore, to aim for establishing a theoretical framework which contains
broader scope, incorporating dipole symmetries and other types of symmetries in a unified way, 
we explore the interplay between the dipole symmetries and higher form symmetries.
To this end, we deal with two types of symmetries on the same footing by 
introducing \textit{higher form dipole symmetries}, and study gauge theories
associated with them. \par
We argue that the algebraic relation between dipole and global charges naturally yields foliated BF theories~\cite{foliated1,foliated2}, BF theories defined on layers of submanifolds, which were studied in the context of fracton physics. 
There has been much progress particularly in $(2+1)$-dimensions [$(2+1)$d], such as relation between foliated BF theories and another exotic symmetry, called subsystem symmetry~(i.e., symmetry on a submanifold)~\cite{PhysRevB.66.054526,seiberg2021exotic,Ohmori:2022rzz,Shimamura:2024kwf}.
Although a few foliated BF theories with dipole symmetries have been studied in $(2+1)$ dimensions~\cite{2023foliated}, theories defined in higher dimensions (i.e., $d>2$) have been much less explored.
  In this work, we introduce foliated BF theories with $p$-form dipole symmetries in any dimension, which host dipole of \textit{spatially extended excitations} (e.g., membranes when $p=2$) and study their physical properties and 't Hooft anomalies of the dipole symmetries.
We find that due to the algebraic relation between global and dipole charges, gauge invariant loops have unusual form, containing linear function of the spatial coordinate. 
This feature leads to the position dependent braiding statistics and unusual GSD dependence on the system size, i.e., the greatest common divisor between the charge~$N$ and the system size. 
Note that such GSD dependence is the manifestation of the UV/IR mixing, stemming from the conservation of the dipole. 
The UV/IR mixing that our theories exhibit is distinct from the one found in previous field theories with subsystem symmetry, such as~\cite{seiberg2021exotic} where the GSD becomes subextensive.
\par
We also show that the theories exhibit a mixed \tht~anomaly between $p$-form and $(d-p)$-form dipole symmetries, which is canceled by an invertible theory defined in one dimensional higher via anomaly inflow mechanism~\cite{CALLAN1985427}.
 The bulk theory has a similar form as the one which was introduced in the study of the anomaly inflow when gauging $1$-form symmetries in $(2+1)$d BF theory~\cite{gaiotto2015generalized}, yet crucial difference from the previous cases is that we need to appropriately impose the flatness condition of the gauge fields as they are associated with the dipole symmetries.
 Our consideration can also be applied to the cases with foliated BF theories with subsystem symmetries~\cite{foliated1,foliated2,Cao:2023rrb}, whose in-depth discussion is provided in appendix.\footnote{%
 See also e.g.,~\cite{Burnell:2021reh,Yamaguchi:2021xeq,Honda:2022shd,okuda2024anomaly} for discussion on anomaly inflow for subsystem symmetries in other theories, and~\cite{lam2024topological} for anomaly inflow for a $(1+1)d$ scalar theory with the second order spatial derivatives based on the so-called wire construction.} 
 This paper would contribute to not only better understanding of fracton physics, but also to achieving the ultimate goal to construct a theoretical framework, incorporating various kinds of symmetries.
 \par
The rest of this paper is structured as follows. 
In Sec.~\ref{sec2}, we introduce algebraic relations between global charges and dipole moments, which play a crucial role throughout this work. 
In Sec.~\ref{sec3}, we introduce BF theories which respect higher-form dipole symmetries. 
We also discuss their physical properties, such as braiding statistics between fractional excitations and the GSD on a torus. 
In Sec.~\ref{sec4}, we study the 't Hooft anomaly and its anomaly inflow for the higher form dipole symmetries. 
Finally, in Sec.~\ref{sec5} we conclude our work with a few remarks. 
Some technical details, including construction of the UV spin models that correspond to the dipole BF theories, and thorough analysis on the BF theories with subsystem symmetry and their anomaly, are provided in Appendixes.

\section{Dipole algebra}\label{sec2}
To systematically introduce the foliated BF theories, we start with introducing \textit{multipole algebra}, which describes the multipole symmetries. 
In the present work, we focus on dipole algebra, which is the simplest algebraic relations between global charges and dipole moments. 
Note that the (zero-form)~dipole algebra was introduced in the previous studies~\cite{hirono2022symmetry,2023foliated}. 
In what follows, we are going to generalize such an algebra to higher form, which is necessary to investigate the BF theories and their anomaly.
\par
Suppose that we have a theory in $(d+1)$-dimensions with conserved $(p-1)$-form charges associated with \textit{global} U(1) and \textit{dipole} symmetries, defined on $(d+1-p)$-dimensional spatial submanifold,~$\Sigma_{d+1-p}$.\footnote{%
Here we take $1\leq p\leq d$. The $p=1$ case corresponds to the ordinary global symmetry.}
We denote the global charge by $Q[\Sigma_{d+1-p}]$ and dipole charge by $Q_{I}[\Sigma_{d+1-p}]$, where the index~$I =1,\cdots ,d$ denotes the dipole degrees of freedom in the~$I$-th spatial direction.
 Also, throughout this work, we interchangeably represent the spatial direction $I=1,2,3,\cdots,$ as $I=x,y,z,\cdots,$~depending on the context.\par
While the charge $Q$ follows the relation
\begin{equation}
    [iP_I,Q]=0,\label{df}
\end{equation}
the dipole charges are subject to the following relation:
\begin{equation}
    [iP_I,Q_J]=\delta_{IJ}Q.\label{eq:re2}
\end{equation}
An intuition behind this relation can be understood by associating the global and dipole charges with $\rho$, $x_J\rho$, where~$\rho$ denotes the density of the U(1) charge, and $x_J$ as the $J$-th spatial coordinate, respectively, and thinking of shifting them by a constant in the $I$-th direction~($I,J=1,\cdots,d$)~\cite{2023foliated}. 
For instance, if we shift the dipole moment~$x_I\rho$ by a constant in the $I$-th~direction, then the change of dipole moment under the shift gives $(x_I+\Delta x_I)\rho-x_I\rho=(\Delta x_I)\rho$, where $\Delta x_I$ is constant, corresponding to the nontrivial commutation relation between the transnational operator and the dipole. 
Such an intuitive understanding of the relation~\eqref{eq:re2} will be useful in the discussion on the BF theories presented in the next section. \par
We write the charges $Q$ and $Q_I$ via integral expression using the $p$-form conserved currents as
\begin{equation*}
   Q[\Sigma_{d+1-p}] =\int_{\Sigma_{d+1-p}}*j^{(p)},\quad  Q_I [\Sigma_{d+1-p}]=\int_{\Sigma_{d+1-p}}*K_I^{(p)} .
\end{equation*}
To reproduce the relation~\eqref{eq:re2}, we demand that 
 \begin{equation}
    *K_I^{(p)}=*k_I^{(p)}-x_I*j^{(p)}\label{dd}
\end{equation}
with $k_I\fl p$ being a local (non-conserved) current. One can verify the relation~\eqref{eq:re2} by~\eqref{dd}. 
With the dipole algebra~\eqref{eq:re2} with~\eqref{dd},
we gauge these symmetries. To do so, 
we introduce~$U(1)$ $p$-form gauge fields $a\fl p$, $A\fii p$ with the coupling term being defined by\footnote{%
As discussed in~\cite{2023foliated}, one can regard the gauge group as U(1), taking the fact that quantization condition of the dipole gauge field depends on the length of the dipole into consideration. As we will see soon below, if we define the theory on
a discrete lattice, it can be understood by that the dipole charge is quantized since the lattice spacing becomes a minimal unit of distance. We set such a length to be 1 throughout this work.} 
\begin{equation}
    S_{c}=\int_{V_{d+1}} \left( a\fl p\wedge *j\fl p+\sum_{I=1}^d A\fii p\wedge *k_I\fl p \right) ,
\end{equation}
where $V_{d+1}$ denotes the spacetime.
We need to have a proper gauge transformation in such a way that the condition of the coupling term being
gauge invariant yields the conservation law of the higher form currents. 
It turns out that the following gauge transformation does this job:
\begin{equation}
    a\fl p\to a\fl p+d\Lambda\fl{p-1}+(-1)^{p-1}\sum_I\sigma\fii{p-1} \wde dx_I,\quad A\fii p\to A\fii p+d\sigma\fii{p-1}.\label{gaugetr1}
\end{equation}
Here, $\Lambda\fl{p-1}$ and $\sigma\fii{p-1}$ denote the $(p-1)$-form gauge parameters.
Indeed, one can verify that the gauge invariance of the coupling term $ S_{c}$ under the gauge transformation~\eqref{gaugetr1} yields
\begin{equation*}
   d*j\fl p=0,\quad d(*k_I\fl p-x_I*j\fl p)=d*K_I\fl p=0.
\end{equation*}
In what follows, following the terminology in fracton physics~\cite{foliated1,foliated2}, we interpret  $dx_I$ as $1$-form field, that we dub \textit{foliation field}:
\begin{equation}
e^I \vcentcolon = dx_I .
\end{equation}
In the context of fracton topological phases, such a field was introduced so that along the foliation field, layers of submanifolds are stacked.
\par
For later purposes, we also define the dipole algebra with hierarchy structure being inverted, which we call \textit{dual dipole algebra}. 
To do this, instead of~\eqref{df} and \eqref{eq:re2}, we think of $d$ global charges $\widehat{Q}_I$ and one dipole charge $\widehat{Q}$ with relation
\begin{equation}
    [iP_I,\widehat{Q}_J]=0,\quad[iP_I,\widehat{Q}]=-\widehat{Q}_I.\label{e}
\end{equation}
Analogous to the argument below~\eqref{eq:re2},
this relation can be intuitively understood by interpreting the dipole and $d$ global charges as $\widehat{\eta}=-\sum_{I=1}^dx_I\rho_I$, and $\rho_I$, respectively, and thinking of the translation operator that acts on it. 
For example, by shifting the dipole~$\widehat{\eta}$ in the $I$-th direction, one obtains the second relation in~\eqref{e}.
Following the similar argument presented around \eqref{dd}-\eqref{gaugetr1}, we define gauge fields associated with the global and dipole charges~\eqref{e} as $\widehat{a}\fii p$ and $\widehat{A}\fl p$ respectively  with the following gauge transformations:
\begin{eqnarray}
    \widehat{a}\fii p\to  \widehat{a}\fii p+d\widehat{\chi}\fii{p-1}-(-1)^{p-1}\widehat{\sigma}\fl{p-1} \wde e^I,\quad\widehat{A}\fl {p}\to\widehat{A}\fl{p}+d\widehat{\sigma}\fl{p-1}.\label{ddd0}
\end{eqnarray}
Here, $\widehat{\chi}\fii{p-1}$ and $\widehat{\sigma}\fl{p-1}$ the are the $(p-1)$-form gauge parameters.\par
The dipole algebra \eqref{eq:re2} is related to the dual one \eqref{e} by inverting the hierarchy structure of the algebra. 
Stated symbolically, 
\begin{equation}
    \begin{Bmatrix}
Q_1 & Q_2 & \cdots&Q_d\\
& Q &&
\end{Bmatrix}
\quad \leftrightarrow\quad 
    \begin{Bmatrix}
    & \widehat{Q} &&\\
\widehat{Q}_1 & \widehat{Q}_2 & \cdots&\widehat{Q}_d
\end{Bmatrix}.\label{triangle}
\end{equation}
These algebras put different mobility constraints on charges. 
In the case of the dipole algebra~\eqref{eq:re2}, a single charge is immobile as dipole moment is conserved in any spatial direction. 
On the contrary, in the case of the dual dipole algebra~\eqref{e}, which consists of $d$ charges (labeled by $I=1,\cdots,d$) and one dipole, the $I$-th charge, $\widehat{Q}_I$, is mobile in the direction perpendicular to the $I$-th direction, yet it is immobile in the $I$-th direction. 
Such mobility constraints play a crucial role in understanding physics of the dipole BF theory that we study in the next section. 
In what follows, we will discuss 't Hooft anomalies in the BF theories, relegating the discussion on the subsystem BF theories to Appendix.~\ref{ap3}.

\section{Dipole BF theories}
\label{sec3}
In this section, we introduce BF theories with the dipole symmetries that we dub dipole BF theories. 
Due to the UV/IR mixing, treatment of the dipole BF theories in the continuum limit is a bit subtle. 
To handle this issue, we put our theory in more appropriate format, that is sometimes called integer BF theory~\cite{PhysRevB.106.045112}, consisting of integer valued gauge fields on a discrete lattice. 
We discuss several physical properties of the model, such as braiding statistics and GSD on torus geometry before diving into the anomaly inflow of the higher dipole symmetries. 

\subsection{Model}
Using the gauge fields $a\fl p$ and $A\fii p$ in~\eqref{gaugetr1} associated with the dipole algebra~\eqref{eq:re2}, we introduce the following gauge invariant $(p+1)$-form fluxes:
\begin{equation}
    f\fl{p+1}\vcentcolon=da\fl p+(-1)^p\sum_IA\fii p\wedge e^I,\quad F\fii{p+1}\vcentcolon=dA\fii p.\label{fl}
\end{equation}
In terms of these fluxes, let us consider 
the following $(d+1)$-dimensional theory, which we call $p$-form dipole BF theory:\footnote{%
Here we take $1\leq p\leq d-1$. In the case of $p=d$, the theory describes a spontaneous symmetry braking phase, which we do not consider in this work.}
\begin{eqnarray}
\mathcal{L} 
&=& \frac{N}{2\pi}\left[b\fl{d-p}\wedge f\fl{p+1} +\sum_I c\fii{d-p}\wedge F\fii{p+1}\right]\nonumber\\
   & =&
     \frac{N}{2\pi}\left[b\fl{d-p}\wde \left(da\fl{p}+(-1)^p\sum_IA\fii{p}\wde e^I\right)+\sum_Ic\fii{d-p}\wde dA\fii{p}\right] ,
\label{pform} 
\end{eqnarray}
where $b\fl{d-p}$ and~$c\fii{d-p}$ denote the $(d-p)$-form gauge fields.
To study physical properties of this theory, such as braiding statistics of fractional excitations and the GSD on a torus geometry, we face an issue with the UV/IR mixing --- the theory is sensitive to the UV physics, making analysis on the theory in the continuum limit much more challenging. 
To remedy this problem, we make use of argument presented in e.g,~\cite{Gorantla:2021svj,2023foliated}, where we introduce \textit{integer BF theories}, consisting of integer valued gauge fields which are defined on a discrete lattice. In this formalism, we discuss braiding statistics and the GSD to 
get more physical insights from the theory~\eqref{pform}
before studying the anomaly inflow.

\subsubsection{Notations}
To do this, we prepare several notations. 
In order to investigate the integer BF theory with dipole symmetry, comprised of integer valued gauge fields, we define an integer valued $p$-form field, located on a~$p$-cell
in a $(d+1)$-dimensional discrete lattice as\footnote{The field with a tilde on the top $\Tilde{(\cdot)}$ represents integer valued field throughout this work.}
\begin{equation}
    \Tilde{A}\fl{p}_{[\mu_1\cdots\mu_p]}(\hat{\mathbf{r}}_A)\vcentcolon=\frac{1}{p!}\sum_\sigma Sgn(\sigma)\Tilde{A}\fl p_{\mu_{\sigma(1)}\cdots\mu_{\sigma(p)}}(\hat{\mathbf{r}}_A),\label{field}
\end{equation}
where $[\mu_1\cdots\mu_p]$ denotes $p$-th anti-symmetrized index and $\sigma$ represents the permutation with its signature denoted by $Sgn(\sigma)$. 
Also, we have introduced the coordinate of the field as $\hat{\mathbf{r}}_A=(\hx_0,\hx_1,\cdots,\hx_d)$.
Without bringing any confusion, we omit the lower case indices and the coordinate of the gauge field on the left hand side of~\eqref{field}, and simply write it as $\Tilde{A}\fl p$.
We introduce two more notations. We define a
differential operator $\Delta$, mapping a field $\Tilde{A}\fl p$ to $\Delta\Tilde{A}\fl p$ on $(p+1)$-cell, which is given by the oriented sum of $\Tilde{A}\fl p$ on the boundary of the $(p+1)$-cell. More explicitly, it is defined as
\begin{equation}
  \Delta \Tilde{A}\fl{p}_{[\mu_1\cdots\mu_{p+1}]}\vcentcolon=
  \frac{1}{p!}\sum_\sigma Sgn (\sigma)\Delta_{\mu_{\sigma(1)}}\Tilde{A}\fl p_{\mu_{\sigma(2)}\cdots\mu_{\sigma(p+1)}},
\end{equation}
which is abbreviated as $\Delta\Tilde{A}\fl p$.
We also define the following integer valued field located on $(p+1)$-cell, which corresponds to the wedge product between a $p$-form field and a foliation field $e^I$ in the continuum:
\begin{equation}
    \Tilde{A}\fl pe^I\vcentcolon=\frac{1}{p!}\sum_\sigma Sgn(\sigma)\Tilde{A}\fl p_{\mu_{\sigma(1)}\cdots\mu_{\sigma(p)}}\delta^I_{\mu_{\sigma(p+1)}}\quad(I=1,\cdots, d).
\end{equation}
Here $\delta^I_J$ represents the Kronecker delta.

\subsubsection{Integer BF theory}
With these preparations, now we are in a good place to discuss the integer BF theory.
The integer BF theory that corresponds to~\eqref{pform}, which is defined on a $(d+1)$-dimensional discrete lattice, reads
\begin{equation}
  \boxed{ 
\mathcal{L}  =\frac{2\pi}{N}\sum_{(p+1)-{\rm cell}}\left[ \tilde{b}\fl{d-p}\left(\Delta\tilde{a}\fl p+(-1)^p\sum_I\Tilde{A}\fii pe^I\right)+\sum_I \Tilde{c}\fl{d-p}\Delta \Tilde{A}\fii p\right].}\label{inbf}
\end{equation}
Here, the gauge fields $\ta\fl p$ and $\tl{A}\fii p$ reside on $p$-cells in the lattice whereas $\tb\fl{d-p}$ and $\tl{c}\fii{d-p}$ do on~dual~$(d-p)$-cells. 
The theory~\eqref{inbf} consists of $d+1$ layers of the $p$-form $\mathbb{Z}_N$~BF theories, corresponding to the first and third terms, and the coupling between the BF theories described by the second term. In this sense, \eqref{inbf} is the~$p$-form analog of the foliated BF theories~\cite{foliated1,foliated2}.
Note that while in the preexisting foliated BF theories are made of subextensive number of layers of the BF theories, our theory consists of the finite number of layers. As shown below, the coupling between the layers brings unusual physical properties such as braiding statistics and the GSD.\par
The theory~\eqref{inbf} admits the following gauge symmetry:
\begin{eqnarray}
&&    \Tilde{a}\fl p\to \Tilde{a}\fl p+\Delta\Tilde{\chi}\fl{p-1}+(-1)^{p-1}\sum_I\Tilde{\sigma}\fii{p-1}e^I+N\Tilde{k}\fl p_a,\quad \Tilde{A}\fii p\to \Tilde{A}\fii p+\Delta\sigma\fii{p-1}+N\Tilde{k}_{A}\fii{p},\nonumber\\
&&  \Tilde{b}\fl{d-p}\to \Tilde{b}\fl{d-p}+\Delta\Tilde{\eta}\fl{d-p}+N\Tilde{k}\fl{d-p}_b,\quad \Tilde{c}\fii{d-p}\to\Tilde{c}\fii{d-p}+\Delta\Tilde{\gamma}\fii{d-p}-(-1)^{d-p}\Tilde{\eta}\fl{d-p}e^I+N\Tilde{k}\fii{d-p}_c. \nonumber\\
    \label{gtr3}
\end{eqnarray}
The integer valued gauge parameters $\tl{\chi}\fl{p-1}$ and $\tilde{\sigma}\fii{p-1}$ are defined on $(p-1)$-cells whereas $\tl{\eta}\fl{d-p-1}$ and $\tl{\gamma}\fii{d-p-1}$ are introduced on dual $(d-p-1)$-cells. Also, $\tl{k}_a\fl p$ and $\tl{k}_A\fii p$ [$\tl{k}_b\fl{d-p}$ and $\tl{k}_c\fii{d-p}$] are integer valued fields on $p$-cell [dual $(d-p)$-cell].
Note that while gauge fields $\Tilde{a}\fl p$ and $\Tilde{A}\fii p$ are subject to the gauge transformation of the dipole algebra~\eqref{gaugetr1}, the other gauge fields $\Tilde{b}\fl{d-p}$ and $\Tilde{c}\fii{d-p}$ follow the dual dipole algebra~\eqref{ddd0}. 
In the following, to get better handle on the theory~\eqref{inbf}, we study several physical properties of the model before discussing the anomaly.
\subsection{Example: $d=3$, $p=1$}
\label{sec32}
We demonstrate unusual properties of the braiding statistics and the GSD of the model~\eqref{inbf} in a specific case, setting $d=3$ and $p=1$.
Argument presented in this subsection can be generalized to any dimension and higher form symmetries, which is given in Appendix~\ref{ap:ge}. \par

\subsubsection{Loops and braiding statistics}
We study braiding statistics between gauge invariant loops of the theory~\eqref{inbf} with $d=3$ and $p=1$:
\[
\mathcal{L}  =\frac{2\pi}{N}\sum_{2-{\rm cell}}\left[ 
\tilde{b}\fl{2}\left(\Delta\tilde{a}\fl{1} -\sum_I\Tilde{A}\fii{1}e^I\right)+\sum_I \Tilde{c}\fl{2}\Delta \Tilde{A}\fii{1}\right].
\]
Analogous to the foliated BF theories of fracton topological phases~\cite{foliated1,foliated2}, the coupling between the BF theories modify the gauge transformations, giving rise to mobility constraints on fractional charges. 
From the gauge transformations~\eqref{gtr3}:
\begin{eqnarray*}
&&    \Tilde{a}\fl{1} \to \Tilde{a}\fl{1} +\Delta\Tilde{\chi}\fl{0} 
+\sum_I\Tilde{\sigma}\fii{0}e^I+N\Tilde{k}\fl{1}_a,\quad 
\Tilde{A}\fii{1} \to \Tilde{A}\fii{1} +\Delta\sigma\fii{0} +N\Tilde{k}_{A}\fii{1},\nonumber\\
&&  \Tilde{b}\fl{2}\to \Tilde{b}\fl{2}+\Delta\Tilde{\eta}\fl{2}+N\Tilde{k}\fl{2}_b,\quad 
\Tilde{c}\fii{2}\to\Tilde{c}\fii{2}+\Delta\Tilde{\gamma}\fii{2}-\Tilde{\eta}\fl{2}e^I+N\Tilde{k}\fii{2}_c ,
\end{eqnarray*}
we can construct several types of gauge invariant operators.
First we have the gauge invariant Wilson loops in the spatial direction, described by
\begin{eqnarray}
    W_a(C )\vcentcolon=\exp\left[\frac{2\pi i}{N}\sum_{C}\left(\ta\fl 1+\sum_I\hx_I\tl{A}\fii 1\right)\right],\quad 
    W^I_A(C )\vcentcolon=\exp\left[\frac{2\pi i}{N}\sum_{C}\tl{A}\fii 1\right],\quad(I=x,y,z) , \label{a}
\end{eqnarray}
where $C$ denotes closed path in subspace of $1$-cells.
We also have gauge invariant surface operators given by
\begin{eqnarray}
 V_b(S^*)\vcentcolon=\exp\left[\frac{2\pi i}{N}\sum_{S^* }\tl{b}\fl 2\right],\quad
 V^I_c(S^{*})\vcentcolon  =
\begin{cases}
   \exp\left[\frac{2\pi i}{N}\sum_{S^{*}}\tc\fii 2\right] & {\rm for}\ S^{*}\perp I\text{-th direction} \\
     \exp\left[\frac{2\pi i}{N}\sum_{S^{*}}\left(\tc\fii 2-\hx_I\tb\fl 2\right)\right] & {\rm for}\ S^{*} \not\perp I\text{-th direction} 
    \end{cases} . \label{c}
\end{eqnarray}
where $S^*$ is closed surface defined in subspace of dual $2$-cells. 
Note that
from the gauge transformation of~$\tc\fii 2$~\eqref{gtr3}, depending on the direction, there are two types of closed surface operators made of $\tc^I$: 
(i) when the surface $S^{*}$
is 
perpendicular to the~$I$-th direction (i.e., the surface is formed at fixed $\hx_I$), we have the gauge invariant surface operator
comprised only of the gauge field~$\tc\fii 2$, corresponding to the first line in~\eqref{c}. 
 (ii) when the surface $S^{*}$ 
is
not perpendicular to the $I$-th direction, composite of $\tc\fii 2$ and $\tb\fl  2$ is the appropriate gauge invariant surface operator,
corresponding to the second line in~\eqref{c}.
\par

While $1$-form and $2$-form gauge invariant loops are also found in the conventional $(3+1)$d BF theories (or the $(3+1)$d toric code~\cite{KITAEV20032,PhysRevB.72.035307} on the lattice), in the present case,
some of the gauge invariant extended operators
 contain linear function of the spatial coordinate~$\hx_I$, which is a unique feature of the dipole BF theories~\cite{ebisu2209anisotropic,2024multipole}.
The form of the loops~\eqref{a} and \eqref{c} can be comprehended by mobility constraints on quasiparticles with the algebraic relations~\eqref{eq:re2} and \eqref{e}.\footnote{ 
The mobility constraints on charges in a system which preserves the dipole moment are also discussed in~e.g.,~\cite{tensor_gauge}.}
We recall the fact that the gauge fields~$\ta\fl1$ and $\tl{A}\fii1$ are associated with the dipole algebra~\eqref{eq:re2}, corresponding to conservation of three dipole moments, forming in $x$, $y$, and $z$-direction, and one global charge. 
Since the dipoles are conserved, a single charge is immobile in any direction whereas the dipole moments are free to move. Correspondingly, a single charge associated with the gauge field~$\ta\fl1$, is not mobile in any of spatial direction. 
To make it mobile, the charge must accompany with the dipole moment, associated with the gauge field~$\tl{A}\fii1$. Also, the dipole moments are free to move -- they are consistent with the form of~\eqref{a}, that is, the form of the Wilson loop of the gauge field~$\ta\fl{1}$ is accompanied with~$\tl{A}\fii1$, and that of the gauge field~$\tl{A}\fii1$ may form in any direction. 
Analogous line of thoughts shows that the form of the loops~\eqref{c} can be understood by the fact that the gauge fields~$\tc\fii2$ and $\tb\fl2$ follow the dual dipole algebra~\eqref{e}: three global charges $\rho_i\:(i=1,2,3)$ and one dipole moment~$-(x\rho_1+y\rho_2+z\rho_3)$. 
For instance, a single charge corresponding to gauge field~$\tc^{x(2)}$, is mobile in the $y$~and $z$-direction and is not in the $x$-direction. To make it mobile in the $x$-direction, the charge must accompany with the charge associated with the dipole gauge field~$\tb\fl2$, which is consistent with~\eqref{c}.
\par
The loops~\eqref{a} and \eqref{c} have several unusual properties compared with the ones in the conventional BF theories. 
To see this, we introduce a lattice translation operator $T_{\hx_I}$ which translates fields
by a unit of the lattice spacing in the $I$-th direction. 
Acting the translation operator on the Wilson loop, we find
\begin{eqnarray}
    \frac{T_{\hx_I}W_a(C )T^{-1}_{\hx_I}}{W_a( T_{\hx_I} (C))}=W_A^I( T_{\hx_I} (C) ),   
\end{eqnarray}
where $T_{\hx_I} (C)$ is a contour obtained by shifting $C$ by one lattice spacing in the $I$-th direction.
Similarly, the surface operator not perpendicular to the $I$-th direction satisfies
\begin{eqnarray}
 \frac{T_{\hx_I}V^I_c(S^*)T^{-1}_{\hx_I}}{V^I_c( T_{\hx_I} (S^* ) )}=
        V_b( T_{\hx_I} (S^* ) )\quad (S^{*} \not\perp I\text{-th direction}).
\end{eqnarray}
where $T_{\hx_I} (S^\ast )$ is a surface obtained by shifting $S^\ast$ by one lattice spacing in the $I$-th direction.
These relations remind us of the dual dipole algebra~\eqref{e} and dipole algebra~\eqref{eq:re2} that we have seen in the previous argument of the dipole symmetry\footnote{%
See also~\cite{2023foliated} for the related relations found in lattice spin models with dipole symmetry.
}, namely, translating a loop by a constant in the $I$-th direction yields another type of loop, which is in line with the dipole and dual dipole algebra where translating a dipole moment gives rise to a charge. 
\par
We also discuss braiding statistics of the loops~\eqref{a} and \eqref{c}. Similar to the conventional BF theories, 
we think of a torus geometry and study the braiding statistics between the loops that wind around the torus in the spatial directions. To this end, we impose the periodic boundary conditions on the lattice and set the system size as $L_{x}\times L_y\times L_z$. Also, we focus on the 
braiding the $1$-form loops that go around the torus in the $x$-direction and $2$-form loops that run along $yz$-plane. Braiding of the loops in the other spatial directions can be analogously discussed. 
The noncontratcible loops of the gauge fields~$\ta\fl p$ and~$\tl{A}\fii p$ that wind around the torus in the $x$-direction are given by
\begin{eqnarray}
&& W_{a:x}(\hy,\hz) :=  W_{a}(C_{x}) =\exp\left[\frac{2\pi i}{N}\alpha_x\sum_{C_{x}}\left(\ta\fl 1+\hx\tl{A}^x+\hy\tl{A}^y+\hz\tl{A}^z\right)\right],\nonumber \\
&& W^I_{A:x}(\hy,\hz) :=  W^I_{A}(C_{x}) =\exp\left[\frac{2\pi i}{N}\sum_{C_{x}}\tl{A}\fii1\right] ,
   \label{ap}
\end{eqnarray}
where $C_{x}$ represents noncontractible paths around the torus in the $x$-direction, 
and
\begin{equation}
\alpha_I \vcentcolon=\frac{N}{\gcd(N,L_I)} \quad (I=x,y,z ) ,
\end{equation}
with $\gcd$ standing for the greatest common divisor. 
An intuition behind the first loop~\eqref{ap} is that the argument of the exponent has linear function of the spatial coordinate~$\hx$, hence the loop has to wind around the torus \textit{multiple times} in order for it to be consistent with the periodic boundary condition.\footnote{%
See e.g., \cite{ebisu2209anisotropic,2023foliated} for the relevant discussion in a different topological model with dipole symmetry.}
The noncontractible loops of the gauge fields~$\tb\fl{d-p}$ and $\tc\fii{d-p}$ that are defined in dual $2$-cells on 
a $yz$-plane are described by
\begin{eqnarray}
&&    V_{b:yz}\left(\hx+1/2\right) := V_{b}(S^*_{yz}) =\exp\left[\frac{2\pi i}{N}\sum_{S^*_{yz}}\tl{b}\fl2\right],\nonumber\\
&&    V^I_{c:yz}\left(\hx+1/2\right) := V^I_{c}(S^*_{yz}) =
    \begin{cases}
        \exp\left[\frac{2\pi i}{N}\sum_{S^*_{yz}}\tl{c}\fii2\right] & {\rm for}\ I=x\\
        \exp\left[\frac{2\pi i}{N}\alpha_I\sum_{S^*_{yz}}\left(\tl{c}\fii2-\hx_I\tb\fl2\right)\right] & {\rm for}\ I=y,z,
    \end{cases}\label{apc}
\end{eqnarray}
where
$S^*_{yz}$ denotes surfaces on $yz$-plane of the dual $2$-cells that go around the torus, forming noncontractible $2$-form loops.
Similar to the first term in~\eqref{ap}, we multiply the integer~$\alpha_I$ with the argument of the exponent of the loop~$V^I_{c:yz}$ with $I=y,z$ 
to ensure that the loop is compatible with the periodic boundary condition.
\par
The nontrivial braiding statistics between the loops~\eqref{ap} and \eqref{apc} is given by
\begin{eqnarray}
     W_{a:x}(\hy,\hz)  V_{b:yz}\left(\hx+1/2\right)&=&  V_{b:yz}\left(\hx+1/2\right) W_{a:x}(\hy,\hz)\exp\left[-i\frac{2\pi \alpha_x}{N}\right]\nonumber\\
     W^I_{A:x}(\hy,\hz)   V^J_{c:yz}\left(\hx+1/2\right)&=& V^J_{c:yz}\left(\hx+1/2\right)W^I_{A:x}(\hy,\hz)
\exp\left[-i\frac{2\pi }{N}\left(\delta_{I,J}+(1-\delta_{I,J})\alpha_J\right)\right]\nonumber\\
W_{a:x}(\hy,\hz)  V^x_{c:yz}\left(\hx+1/2\right)&=&  V^x_{c:yz}\left(\hx+1/2\right) W_{a:x}(\hy,\hz)\exp\left[-i\frac{2\pi \alpha_x}{N}\hx\right].
     \label{22}
\end{eqnarray}
Due to the fact that some of the loops~\eqref{ap}~\eqref{apc} contain linear function of the spatial coordinate,~$\hx_I$, we have unusual braiding statistics between the loops. 
Especially,  a phase factor obtained by braiding between the loops $V_{c:yz}^x$ and $ W_{a:x}$
depends on the spatial coordinates, as shown in the third line in~\eqref{22}. 
We regard it as the feature specific to the BF theory with the dipole symmetries.\footnote{ Position dependent braiding statistics is also studied in the Maxwell theory with dipole symmetry~in~\cite{hirono2022symmetry} and the Higgsed phase of a tensor gauge theory in~\cite{han2024dipolar}.}

\subsubsection{Ground state degeneracy}
\label{322}
Due to the dipole symmetries, the model exhibits the unusual behaviour of the GSD when we place it on torus geometry. 
To see how the model exhibits the unusual GSD dependence on the system size, we place the theory on a discrete lattice with system size $L_x\times L_y\times L_z$ with periodic boundary conditions. 
The equations of motions of the BF theory~\eqref{inbf}
ensure that local gauge invariant fluxes are trivial. Yet, there are non-local ones forming noncontractible loops of the gauge fields, contributing to the GSD. Focusing on such loops of the gauge fields $\Tilde{a}\fl 1_i$ and $\Tilde{A}\fl 1_i$, we evaluate the number of distinct configurations of loops, which amounts to the GSD on torus. 
The gauge invariant noncontractible loops of $\Tilde{a}\fl 1_i$ and $\Tilde{A}\fl 1_i$ are given by\footnote{Here we explicitly write the components of the gauge fields for the sake of illustration.}
\begin{eqnarray}
&&    W_{a:x}(\hy,\hz)=\exp\left[\frac{2\pi i}{N}\alpha_x\sum_{\hx=1}^{L_x}\left(\ta_x\fl 1+\sum_I\hx_I\tl{A}_x\fii 1\right)\right],\quad   
W_{a:y}(\hx,\hz)=\exp\left[\frac{2\pi i}{N}\alpha_y\sum_{\hy=1}^{L_y}\left(\ta_y\fl 1+\sum_I\hx_I\tl{A}_y\fii 1\right)\right],\nonumber\\
&&   W_{a:z}(\hx,\hy)=\exp\left[\frac{2\pi i}{N}\alpha_z\sum_{\hz=1}^{L_z}\left(\ta_z\fl 1+\sum_I\hx_I\tl{A}_z\fii 1\right)\right],\label{lp1}
\end{eqnarray}
\begin{eqnarray}
    W^I_{A:x}(\hy,\hz)=\exp\left[\frac{2\pi i}{N}\sum_{\hx=1}^{L_x}\tl{A}\fii1_x\right],\quad  W^I_{A:y}=\exp\left[\frac{2\pi i}{N}\sum_{\hy=1}^{L_y}\tl{A}\fii1_y\right],\quad  W^I_{A:z}=\exp\left[\frac{2\pi i}{N}\sum_{\hz=1}^{L_z}\tl{A}\fii1_z\right].\label{lp2}
\end{eqnarray}
Note that the loops~$W_{a:x}(\hy,\hz)$ and $W_{A:x}^I(\hy,\hz)$ have already appeared in the previous discussion on the braiding statistics~\eqref{ap}.
To properly count the number of distinct noncontractible loops of the gauge fields, we need to check several constraints imposed on the loops. 
Indeed, from~\eqref{44}, it follows that the loop~$W_{a:x}(\hy,\hz)$ is deformable in the $y$- and $z$-directions, indicating that the loop does not depend on the coordinate $(\hy,\hz)$. 
By the same token, one can show that the other loops in~\eqref{lp1} and \eqref{lp2} do not depend on the coordinates. 
This stems from the fact that 
\begin{eqnarray}
    \Delta_i\left(\Tilde{a}_j\fl 1+\sum_I\hat{x}_IA\fii 1_j\right)- \Delta_j\left(\Tilde{a}_i\fl 1+\sum_I\hat{x}_IA\fii 1_i\right)=0 \quad (i,j=x,y,z ),
    \label{44}
\end{eqnarray}
where $\hat{x}_I$ represents the $I$-th spatial coordinate of the lattice. 
This relation is verified from the equations of motions for $\tb\fl 2$ and $\tc\fii 2$ in the BF theory~\eqref{inbf}: 
\begin{eqnarray}
      \Delta_i\tilde{a}\fl 1_j-\Delta_j\Tilde{a}\fl 1_i-\Tilde{A}^{j(1)}_i+\Tilde{A}^{i(1)}_j=0,\quad\Delta_i\Tilde{A}\fii 1_j-\Delta_j\Tilde{A}\fii 1_i=0 .
 \end{eqnarray}
Indeed, we rewrite the left hand side of~\eqref{44} as
\begin{eqnarray}
&&  \left(\Delta_i\Tilde{a}_j\fl 1+\sum_I\Delta_i(\hat{x}_I)A\fii 1_j\right)+\sum_I\hx_I\Delta_i\Tilde{A}\fii 1_j- \left(\Delta_j\Tilde{a}_i\fl 1+\sum_I\Delta_j(\hat{x}_I)A\fii 1_i\right)-\sum_I\hx_I\Delta_j\Tilde{A}\fii 1_i\nonumber\\
&=&  \Delta_i\tilde{a}\fl 1_j-\Delta_j\Tilde{a}\fl 1_i-\Tilde{A}^{j(1)}_i+\Tilde{A}^{i(1)}_j+\sum_I\hx_I(\Delta_i\Tilde{A}\fii 1_j-\Delta_j\Tilde{A}\fii 1_i)=0.
\end{eqnarray}
Here we have used the fact that $\Delta_i(\hx_I)=\delta^i_I$.
\par
Also, due to the periodic boundary condition, we have
\begin{eqnarray}
     W_{a:x}(\hy+L_y,\hz)= W_{a:x}(\hy,\hz),\quad  W_{a:y}(\hx+L_x,\hz)= W_{a:y}(\hx,\hz),
\end{eqnarray}
which is rewritten as
\begin{eqnarray}
    (W^y_{A:x})^{\alpha_xL_y}= (W^x_{A:y})^{\alpha_yL_x}=1.\label{35}
\end{eqnarray}
Furthermore, setting $i=x$ and $j=y$, and summing over $\hx$ and $\hy$
in~\eqref{44} yields
\begin{eqnarray*}
    L_x\sum_{\hy=1}^{L_y}\tl{A}_y^x= L_y\sum_{\hx=1}^{L_x}\tl{A}_x^y
\end{eqnarray*}
from which we have
\begin{eqnarray}
    (W^y_{A:x})^{L_x}=  (W^x_{A:y})^{L_y}.\label{36}
\end{eqnarray}
From the two conditions~\eqref{35} and \eqref{36}, it follows that there are $N\times\gcd(N,L_x,L_y)$ distinct loops of $W^y_{A:x}$ and $W^x_{A:y}$.
Analogous line of thought shows that there are $N\times\gcd(N,L_y,L_z)$ [resp.~$N\times\gcd(N,L_x,L_z)$] distinct loops of $W^y_{A:z}$ and $W^z_{A:y}$~[resp.~$W^x_{A:z}$ and $W^x_{A:z}$]. There is no constraint imposed on the three loops, $W^x_{A:x}$, $W^y_{A:y}$, and $W^z_{A:z}$.
\par
In summary, there are $\prod_{i=x,y,z}\gcd(N,L_i)$ distinct noncontractible loops in~\eqref{lp1} whereas there are 
\begin{eqnarray*}
    \left(N\times\gcd(N,L_x,L_y)\right)\times  \left(N\times\gcd(N,L_y,L_z)\right)\times  \left(N\times\gcd(N,L_x,L_z)\right)\times N^3
\end{eqnarray*}
distinct loops in~\eqref{lp2}. 
Therefore, the GSD, which amounts to the total number of distinct configurations of the noncontractible loops of the gauge fields $\ta\fl1_i$ and $\tl{A}\fii1_i$, is given by
\begin{equation}
    GSD=N^6\times \gcd(N,L_x)\times  \gcd(N,L_y)\times  \gcd(N,L_z)\times \gcd(N,L_x,L_y)\times  \gcd(N,L_y,L_z)\times  \gcd(N,L_z,L_x).
\end{equation}
The GSD dependence on the greatest common divisor between charge $N$ and system size is the manifestation of the UV/IR mixing. Such UV/IR mixing is distinct from the one in previous theories with subsystem symmetry where the GSD becomes subextensive.
\par
By the analogous line of thoughts, one can evaluate the GSD for general $p$ and $d$ on torus geometry.
Relegating the details to Appendix~\ref{ap:ge}, the GSD for the system size $L_{1}\times\cdots\times  L_{d}$ is found to be
\begin{equation}
\boxed{GSD=N^{K(d,p)}\times\prod_{1\leq i_1<i_2\cdots <i_p\leq d}\gcd (N,L_{i_1},L_{i_2},\cdots,L_{i_p})\times\prod_{1\leq i_1<i_2\cdots <i_{p+1}\leq d}\gcd (N,L_{i_1},L_{i_2},\cdots,L_{i_{p+1}})}\label{gs}
\end{equation}
with
\begin{equation}
 K(d,p)\vcentcolon=p\times\binom{d+1}{p+1}.\label{34}
\end{equation}
In Appendix~\ref{app:st}, we demonstrate concrete UV spin models corresponding to the BF theory with dipole symmetry~\eqref{inbf} which exhibit the unusual behavior of the GSD~\eqref{gs}.

\section{Anomaly inflow}
\label{sec4}
In this section, we argue that the dipole BF theory~\eqref{inbf} has a mixed~\tht~anomaly between the higher form dipole symmetries.
We also find a bulk theory in one higher dimensions which cancels the 't Hooft anomaly by anomaly inflow mechanism~\cite{CALLAN1985427}.

\subsection{Review of $1$-form gauging in $(2+1)$d BF theory}
Before discussing anomaly inflow in the foliated BF theories~\eqref{inbf}, we succinctly review how anomaly inflow mechanism works in the case of conventional $(2+1)$d BF theory~\cite{gaiotto2015generalized}. 
Readers who are already familiar with this argument may skip this subsection.\par
We start with the following BF theory
\begin{equation}
\mathcal{L}    =\frac{N}{2\pi}a^{(1)}\wedge db^{(1)}.
\end{equation}
To gauge $1$-form global symmetries, we introduce a background $2$-form field $\beta\fl{2}$ to gauge one of the $1$-form symmetries. 
Then we rewrite the theory as
\begin{equation}
\mathcal{L} =\frac{N}{2\pi}a^{(1)}\wedge (db^{(1)}-\beta\fl{2}) ,
\label{2}
\end{equation}
which respects the $1$-form background gauge symmetry:
\begin{equation}
    b\fl{1}\to b\fl{1}+\lambda\fl{1},\quad \beta\fl{2}\to\beta\fl{2}+d\lambda\fl{1}.\label{3}
\end{equation}
Likewise, one could gauge other $1$-form symmetry by introducing a background $2$-form gauge field~$\alpha\fl{2}$ via
\begin{equation}
\mathcal{L}  =\frac{N}{2\pi}b^{(1)}\wedge (da^{(1)}-\alpha\fl{2})\label{4}
\end{equation}
which has the following background gauge symmetry
\begin{equation}
       a\fl{1}\to a\fl{1}+\sigma\fl{1},\quad \alpha\fl{2}\to\alpha\fl{2}+d\sigma\fl{1}.\label{5}
\end{equation}
Now we gauge both $1$-form symmetries simultaneously, namely
\begin{equation}
\mathcal{L} =\frac{N}{2\pi}a^{(1)}\wedge (db^{(1)}-\beta\fl{2})-\frac{N}{2\pi}b\fl{1}\wedge \alpha\fl{2}.\label{theory}
\end{equation}
However, the theory~\eqref{theory} is not invariant under the gauge transformations~\eqref{3} and \eqref{5}, signaling a 't Hooft anomaly.
Indeed, under the gauge transformations,
we find 
\begin{equation}
\delta\mathcal{L} 
=-\frac{N}{2\pi}
\left[\sigma\fl{1}\wedge\beta\fl{2}+\lambda\fl{1}\wde\alpha\fl{2}+\lambda\fl{1}\wde d\sigma\fl{1} \right].
\label{ano}
\end{equation}
This cannot be removed by adding counter terms consisting of the background gauge fields $\alpha\fl{2}$ and $\beta\fl{2}$ to $\mathcal{L}$.
Therefore \eqref{ano} implies that the theory has the mixed 't Hooft anomaly between the two $1$-form symmetries, which is an obstruction to gauge the both symmetries simultaneously.

It is known that this anomaly can be canceled by combining the theory with an appropriate bulk theory in $(3+1)$ dimensions:
\begin{equation}
    S_{\rm inflow}=\frac{N}{2\pi} \int _{M_{3+1}} \Bigl[ G\fl{1}\wde d\alpha\fl{2}+K\fl{1}\wde d\beta\fl{2} -\alpha\fl{2}\wde \beta\fl{2} \Bigr] .
    \label{inflow}
\end{equation}
Here, we have introduced the $1$-form fields $G\fl{1}$ and $K\fl{1}$ to ensure the flatness condition of the background gauge fields~$\alpha\fl{2}$ and $\beta\fl{2}$. 
To see this inflow term actually cancels the anomaly \eqref{ano}, we first note that 
the theory respects the following background gauge symmetry
\begin{eqnarray}
     \alpha\fl{2}\to\alpha\fl{2}+d\sigma\fl{1},\quad K\fl{1}\to K\fl{1}-\sigma\fl{1},\nonumber\\
     \beta\fl{2}\to\beta\fl{2}+d\lambda\fl{1},\quad G\fl{1}\to G\fl{1}-\lambda\fl{1} , \label{d}
\end{eqnarray}
up to total derivative.
Then, under this transformation,
variant of the theory~\eqref{inflow} is given by
\begin{equation}
    \delta S_{\rm inflow}
    = -\int_{M_{3+1}} d\left[ \sigma\fl{1}\wedge\beta\fl{2}+\lambda\fl{1}\wde\alpha\fl{2}+\lambda\fl{1}\wde d\sigma\fl{1} \right].
\end{equation}
Hence, if we have the theory~\eqref{inflow} with the boundary in the $z$-direction, say at $z=0$, (i.e., the theory is defined in $z\geq 0$), the
mixed 't Hooft anomaly~\eqref{ano} is canceled, which is the well-known anomaly inflow of $1$-form symmetries. Below, we are going to apply the similar logic to our integer dipole BF theories~\eqref{inbf} to discuss the anomaly inflow when gauging the dipole symmetries.

\subsection{Anomaly inflow for dipole symmetries}
Now we turn to anomaly inflow for the dipole BF theory~\eqref{inbf}. 
Introducing integer valued background gauge fields~$\Tilde{\alpha}\fl{p+1}$ and~$\Tilde{\Gamma}\fii{p+1}$, which reside on $(p+1)$-cell,
we first gauge the $p$-form symmetries via
\begin{eqnarray}
\mathcal{L} =\frac{2\pi}{N}\sum_{(p+1)-{\rm cell}}\left[ \tilde{b}\fl{d-p}\left(\Delta\tilde{a}\fl p+(-1)^p\sum_I\Tilde{A}\fii pe^I-\Tilde{\alpha}\fl{p+1}\right)+\sum_I \Tilde{c}\fl{d-p}\left(\Delta \Tilde{A}\fii p-\Tilde{\Gamma}\fii{p+1}\right)\right],\label{inbf2}
\end{eqnarray}
The theory admits the following background gauge symmetry\footnote{%
Throughout subsection, $\tl{\lambda}_*^*$ denotes a gauge parameter.}:
\begin{eqnarray}
&& \ta\fl p\to\ta\fl p+\tl{\lambda}_a\fl p, \quad\tA\fii p\to\tA\fii p+\tl{\lambda}_A\fii p,\nonumber\\
&& \tl{\alpha}\fl{p+1}\to \tl{\alpha}\fl{p+1}+\Delta\tl{\lambda}_a\fl p+(-1)^p\sum_I\tl{\lambda}_A\fii pe^I,\quad\tl{\Gamma}\fii{p+1}\to \tl{\Gamma}\fii{p+1}+\Delta\tl{\lambda}_A\fii p. \label{gauge2}
\end{eqnarray}
One can also gauge the $(d-p)$-form symmetries by introducing the integer valued background gauge fields~$\tl{\beta}\fl{d-p+1}$ and $\tl{\Xi}\fii{d-p+1}$ on dual $(d-p+1)$-cell.
It can be accomplished by
\begin{eqnarray}
\mathcal{L}  &=& \frac{2\pi}{N}\biggl[\sum_{(p+1)-{\rm cell}} \biggl\{\tilde{b}\fl{d-p}\left(\Delta\tilde{a}\fl p+(-1)^p\sum_I\Tilde{A}\fii pe^I\right)+\sum_I \Tilde{c}\fl{d-p}\Delta \Tilde{A}\fii p\biggr\} \nonumber\\
    &&~~~~~~+\sum_{p-{\rm cell}}\biggl\{ -(-1)^{(p+1)(d+1)}\ta\fl p\tl{\beta}\fl{d-p+1}-(-1)^{(p+1)(d+1)}\sum_I\tA\fii p\tl{\Xi}\fii{d-p+1}\biggr\}\biggr].\label{bf44}
\end{eqnarray}
The theory~\eqref{bf44} respects the following gauge symmetry
\begin{eqnarray}
&&  \tb\fl{d-p}\to \tb\fl{d-p}+\tl{\lambda}_b\fl{d-p},\quad \tc\fii{d-p}\to \tc\fii{d-p}+\tl{\lambda}_c\fii{d-p},\nonumber\\
&& \tl{\Xi}\fii{d-p+1}\to \tl\Xi\fii{d-p+1}+\Delta\tl{\lambda}_c\fii{d-p}-(-1)^{d-p}\tl{\lambda}_b\fl{d-p}e^I,\quad \tl{\beta}\fl{d-p+1}\to \tl{\beta}\fl{d-p+1}+\Delta\tl{\lambda}_b\fl{d-p+1}.
    \label{gauge3}
\end{eqnarray}
We gauge the both $p$-form and $(d-p)$-form symmetries by considering
\begin{eqnarray}
\mathcal{L} &=& \frac{2\pi}{N}\biggl[ \sum_{(p+1)-{\rm cell}} \biggl\{\tilde{b}\fl{d-p}\left(\Delta\tilde{a}\fl p+(-1)^p\sum_I\Tilde{A}\fii pe^I-\Tilde{\alpha}\fl{p+1}\right)+\sum_I \Tilde{c}\fl{d-p}\left(\Delta \Tilde{A}\fii p-\Tilde{\Gamma}\fii{p+1}\right)\biggr\} \nonumber\\
 &&~~~~~~+\sum_{p-{\rm cell}}\biggl\{ -(-1)^{(p+1)(d+1)}\ta\fl p\tl{\beta}\fl{d-p+1}-(-1)^{(p+1)(d+1)}\sum_I\tA\fii p\tl{\Xi}\fii{d-p+1}\biggr\}\biggr].\label{inbf3}
\end{eqnarray}
However, the theory~\eqref{inbf3} is not invariant under the background gauge transformations~\eqref{gauge2} and \eqref{gauge3}, signaling the mixed~\tht~anomaly.
Indeed, the variant of the action under the transformations
reads
\begin{eqnarray}
\delta\mathcal{L}  
&=& -\frac{2\pi}{N}\biggl[\sum_{(p+1)-{\rm cell}}\biggl(\tlm{b}\fl{d-p}\tl{\alpha}\fl{p+1}+\sum_I\tlm{c}\fii{d-p} \tl{\Gamma}\fii{p+1}\biggl)  \nonumber\\
&&~~~~~~~~~ +\sum_{p-{\rm cell}}\biggl\{(-1)^{(d+1)(p+1)}\tlm{a}\fl p \left(\tl{\beta}\fl{d-p+1}+\Delta \tlm{b}\fl{d-p}\right)
    \nonumber\\
&&~~~~~~~~~  +(-1)^{(d+1)(p+1)}\sum_I\tlm{A}\fii p\left(\tl{\Xi}\fii{d-p+1}+\Delta\tlm{c}\fii{d-p}-(-1)^{d-p}\tlm{b}\fl{d-p}e^I\right)\biggr\}\biggr].\label{hy}
\end{eqnarray}
This variation cannot vanish even if we add any counter term consisting of the background gauge fields $\tl{\alpha}\fl{p+1}$, $\tl{\Gamma}\fii{p+1}$, $\tl{\beta}\fl{d-p+1}$ and $\tl{\Xi}\fii{d-p+1}$.\footnote{%
Note that cancelling $\delta\mathcal{L}$ needs $(d+1)$-form terms, involving a product of the $(p+1)$-form and $(d-p+1)$-form background gauge fields. However, we cannot construct such terms since they must be $(d+2)$-form or higher.
}
Therefore the dipole BF theory has the mixed 't Hooft anomaly between the $p$-form and $(d-p)$-form symmetries and we cannot gauge them simultaneously.

As in the standard BF theory, this anomaly can be canceled by attaching the dipole BF theory to a bulk theory in one higher dimensions via an extension of anomaly inflow as follows.
Let us consider the following $(d+2)$-dimensional theory 
\begin{eqnarray}
 \mathcal{L}_{\rm inflow} = \mathcal{L}_0 +\mathcal{L}_{\rm spt}, 
\label{re}
\end{eqnarray}
where 
\begin{eqnarray}
\mathcal{L}_0 &=& -\frac{2\pi}{N}\biggl[\sum_{(p+2)-{\rm cell}}\biggl\{\tl{M}
\fl{d-p}\left(\Delta\tl{\alpha}\fl{p+1}+(-1)^{p+1}\tl{\Gamma}\fii{p+1}e^I\right)+\sum_I\tl{N}\fii{d-p}\Delta \tl{\Gamma}\fii{p+1}\biggr\}\nonumber\\
&&~~~~~~~~~ +\sum_{p-\rm cell}\biggl\{\sum_I\tl{O}\fii p\left(\Delta\tl{\Xi}\fii{d-p+1}-(-1)^{d-p+1}\tl{\beta}\fii{d-p+1}e^I \right)+\tl{P}\fl{p}\Delta\tl{\beta}\fl{d-p+1}\biggr\}\biggl],
\end{eqnarray}
and
\begin{equation}
\boxed{
\mathcal{L}_{\rm spt} =-(-1)^{(p+1)(d+1)}\frac{2\pi}{N}\sum_{(p+1)-{\rm cell}}\biggl[\tl{\alpha}\fl{p+1}\tl{\beta}\fl{d-p+1}+\sum_I\tl{\Gamma}\fii{p+1}\tl{\Xi}\fii{d-p+1}\biggl]} .
\label{50}
\end{equation}
An intuition behind 
the bulk term~\eqref{re} is that the term $\mathcal{L}_0$ is what corresponds to the first two terms in~\eqref{inflow} to ensure the flatness condition of the gauge fields, $\Tilde{\alpha}\fl{p+1}$, $\Tilde{\Gamma}\fii{p+1}$, $\Tilde{\beta}\fl{d-p+1}$ and~$\Tilde{\Xi}\fii{d-p+1}$, introducing the four auxiliary fields, $\tl{M}\fl{d-p}$, $\tl{N}\fii{d-p}$, $\tl{O}\fii p$ and $\tl{P}\fii p$. 
Note that the flatness condition has the different form compared with usual cases, as we deal with the dipole gauge fields.
Also, analogous to the last term in~\eqref{inflow}, $\mathcal{L}_{\rm spt}$ describes an invertible phase comprised of the background gauge fields. 
To see how 
the bulk term~\eqref{re} works for cancelling the anomaly, we first note that up to the total derivative,
the theory~\eqref{re} respects the background gauge symmetries in~\eqref{gauge2} and \eqref{gauge3}, jointly with 
\begin{eqnarray}
&&    \tl{M}\fl{d-p}\to  \tl{M}\fl{d-p}+(-1)^{d-p+1}\tlm{b}\fl{d-p},\quad\tl{N}\fii{d-p}\to\tl{N}\fii{d-p}+(-1)^{d-p+1}\tlm{c}\fii{d-p}\nonumber\\
&&    \tl{O}\fii p\to\tl{O}\fii p+(-1)^{d(p+1)}\tlm{A}\fii p,\quad \tl{P}\fl p\to \tl{P}\fl p+(-1)^{d(p+1)}\tlm{a}\fl p.
\end{eqnarray}
Indeed, under these transformations, one has
\begin{eqnarray}
 \delta\mathcal{L}_{\rm inflow}
&=& -\frac{2\pi}{N}\Delta\biggl[\sum_{(p+1)-{\rm cell}}\biggl(\tlm{b}\fl{d-p}\tl{\alpha}\fl{p+1}+\sum_I\tlm{c}\fii{d-p}\tl{\Gamma}\fii {p+1}\biggl)   \nonumber\\
&&~~~~~~~~~~ +\sum_{p-{\rm cell}}\biggl\{(-1)^{(d+1)(p+1)}\tlm{a}\fl p\left(\tl{\beta}\fl{d-p+1}+\Delta \tlm{b}\fl{d-p}\right)    \nonumber\\
&&~~~~~~~~~~~~~~~ +(-1)^{(d+1)(p+1)}\sum_I\tlm{A}\fii p\left(\tl{\Xi}\fii{d-p+1}+\Delta\tlm{c}\fii{d-p}-(-1)^{d-p}\tlm{b}\fl{d-p}e^I\right)\biggr\}\biggr].\nonumber
\end{eqnarray}
The terms inside the large bracket are identical to~\eqref{hy}.

\section{Conclusion}
\label{sec5}
Construction of unified theoretical scheme which incorporates various kinds of symmetries is one of the active area of researches for recent years. 
In this work, we focus on one of the symmetries that has recently emerged in the context of fracton topological phases, multipole symmetries, and incorporate such symmetries into other types of symmetries, higher form symmetries. 
We introduce foliated BF theories consisting of layers of higher form BF theories and couplings between the layers. 
The model admits dipoles of spatially extended excitations, leading to unusual behavior; the position dependent braiding statistics between the loops of the higher form gauge fields, and system size dependence of the GSD. 
We also discuss the 't Hooft anomaly of the foliated higher form BF theory. 
Incapability of gauging higher form dipole symmetries signals the mixed~\tht~anomaly, which is canceled by invertible phases with appropriated terms that ensure the flatness condition of the dipole gauge fields. 
Exploration of the higher form dipole BF theories and their anomaly inflow are central findings in this work, which we believe contribute to not only better understanding of fracton physics, but also to achieving the ultimate goal to construct a theoretical framework that unifies various types of symmetries.
In addition, one can also extend our analysis to the case with the subsystem symmetries, the details of which are given in Appendix.~\ref{ap3}.
\par
Before closing this section, let us make a few comments on future perspectives regarding this work.
In this paper, we have seen the dipole and dual dipole algebra given in~\eqref{eq:re2}, \eqref{e} and \eqref{triangle}. 
Studying the Maxwell theory with such an algebra and especially its dualities would be an interesting direction. 
Relatedly, it could be intriguing to see if we have duality defect lines or even more exotic noninvertible duality defects. 
Investigating whether there is a $\theta$-term in the Maxwell theory could be an another direction to pursue. 
Also, it would be interesting to address whether one can deal with other types of symmetries in the format that we introduced in this paper. 
One candidate would be higher group symmetries, such as~$2$-group~\cite{Cordova:2018cvg}.
When studying theories with dipole symmetries, we sometimes introduce tensor gauge theories, gauge theories with higher order spatial derivatives~\cite{Pretko:2018jbi}. 
It could be interesting to see whether our foliated BF theories~\eqref{inbf} are described by such tensor gauge theories, especially higher form analog of the tensor gauge theories~\cite{pai2018fractonic,shenoy2020k}. 
\par
Investigating whether one can establish a lattice model of the invertible field theory that appears in the discussion on the anomaly inflow~\eqref{50} is an another interesting direction. 
In the case of the standard BF theory in $(2+1)$d (toric code), the mixed anomaly when gauging $1$-form symmetries is canceled by the invertible theory~\eqref{inflow}, which has UV lattice counterpart described by the cluster state (a.k.a. Raussendorf-Bravyi-Harrington cluster state~\cite{rbh_PhysRevA.71.062313}). 
Indeed, one can show that on the boundary of such a state yields the toric code.
One naively wonders a similar cluster state can be constructed which give rise to the topological model corresponding to the dipole BF theory~\eqref{inbf} on the boundary. \par
In this work, we have seen that the \tht~anomaly is canceled by the invertible phases consisting of the dipole gauge fields. 
Investigating whether such invertible phases are of usefulness in classification of symmetry protected topological phases with global dipole symmetries would be an interesting and important direction (see~e.g.~\cite{dSPT} for relevant discussion in the case of $(1+1)$d).
Since our invertible theories are defined in any dimension, our consideration may contribute to classifying such phases. We hopefully leave this issue for future investigations.
\\
\\
\textit{Acknowledgment}--
We thank N. Atlam, B.~Han, J.H.~Han, Q.~Jiang, A.~Nayak, Y.~Oh, S.~Shimamori, K.~Shiozaki,Y.~You, for helpful discussion. We also thank J.H.~Han for sharing draft of his work~\cite{han2024dipolar} with us.
H.~E. is supported by KAKENHI-PROJECT-23H01097.
M.~H. is supported by MEXT Q-LEAP, JSPS Grant-in-Aid for Transformative Research Areas (A) ``Extreme Universe" JP21H05190 [D01], JSPS KAKENHI Grant Number 22H01222, JST PRESTO Grant Number JPMJPR2117.
T.~N. is supported by JST SPRING, Grant Number JPMJSP2110, and RIKEN Junior Research Associate Program.

\appendix
\section{Ground state degeneracy for general $p$ and $d$}\label{ap:ge}
In this appendix, we derive the GSD of the BF theory with the dipole symmetry in the generic case of~$p$ and dimension $d$ given by \eqref{gs}. 
Since discussion presented in this section closely 
parallels the one in the main text~(Sec.~\ref{sec32}), we give the derivation succinctly. 
To start, 
we impose the periodic boundary condition on the lattice and set the system size as $L_{1}\times\cdots\times  L_{d}$. 
The GSD amounts the number of distinct configurations of noncontractible $p$-form loops of the gauge fields $\ta\fl p$ and $\tl{A}\fii p$ that wind in the spatial direction of the torus. 
Defining the indices
\begin{eqnarray*}
    \{i_k~(1\leq k\leq p)|1\leq i_1<i_2<\cdots<i_p\leq d\},\quad  \{j_l~(1\leq l\leq d-p)|1\leq j_1<j_2<\cdots<j_{d-p}\leq d\},
\end{eqnarray*}
where $\{i_k\}$ and $\{j_l\}$ are complement with each other, i.e., $\{i_k\}\cup\{j_l\}=\{1,2,\cdots,d\}$, we find that 
the noncontractible~$p$-form loops which wind around the torus in the 
$i_1,\cdots,i_p$-th spatial directions,
have the following forms:
\begin{eqnarray}
    W_{a;(i_1,\cdots,i_p)}(\hx_{j_1},\cdots,\hx_{j_{d-p}})&=&\exp\left[\frac{2\pi i}{N}\alpha_{(i_1,\cdots,i_p)}\sum_{x_{i_1}=1}^{L_{i_1}}\cdots \sum_{x_{i_p}=1}^{L_{i_p}}\left(\ta_{[i_1i_2\cdots i_p]}\fl p-(-1)^p\sum_I\hx_I\tl{A}_{[i_1i_2\cdots i_p]}\fii p\right)\right],\nonumber\\
    W_{A^I;(i_1,\cdots,i_p)}(\hx_{j_1},\cdots,\hx_{j_{d-p}})&=&\exp\left[\frac{2\pi i}{N}\sum_{x_{i_1}=1}^{L_{i_1}}\cdots \sum_{x_{i_p}=1}^{L_{i_p}}\tl{A}_{[i_1i_2\cdots i_p]}\fii p\right].
    \label{ap:lpa}
\end{eqnarray}
Here, the integer $\alpha_{(i_1,\cdots,i_p)}$ is defined by $\alpha_{(i_1,\cdots,i_p)}\vcentcolon=\frac{N}{\gcd(N,L_{i_1},\cdots,L_{i_p})}$. 
Such an integer is multiplied with the argument in the exponent in the first term of~\eqref{ap:lpa} to ensure that the loop, which contains the linear term of the spatial coordinate, is consistent with the periodic boundary condition~\cite{ebisu2209anisotropic,PhysRevB.107.125154}.\par
There are several constrains on the loops~\eqref{ap:lpa}. From equations of motion for $\tc\fii{d-p}$ in~\eqref{inbf}, one finds that the loop~$W_{A^I;(i_1,\cdots,i_p)}(\hx_{j_1},\cdots,\hx_{j_{d-p}})$ is deformable so that is does not depend on the spatial coordinate~$\hx_{j_l}$.
Due to the periodic boundary conditions on the loop~$W_{a;(i_1,\cdots,i_p)}(\hx_{j_1},\cdots,\hx_{j_{d-p}})$, such as $W_{a;(i_1,\cdots,i_p)}(\hx_{j_1}+L_{i_1},\cdots,\hx_{j_{d-p}})=W_{a;(i_1,\cdots,i_p)}(\hx_{j_1},\cdots,\hx_{j_{d-p}})$, 
we have 
\begin{eqnarray}
   \left( W_{A^I;(i_1,\cdots,i_p)}\right)^{L_{I}\alpha_{(i_1,\cdots,i_p)}}=1\quad(I\neq {i_1},\cdots,i_p).\label{c1}
\end{eqnarray}
From the equation of motion for $\tb\fl{d-p}$ in~\eqref{inbf}, we obtain
\begin{equation}
    \left(W_{A^I;(i_1,\cdots,i_p)}\right)^{L_{I}}\times  \left(W_{A^{i_1};(i_2,\cdots,i_p,I)}\right)^{-L_{i_1}}\times  \left(W_{A^{i_2};(i_2,i_3,\cdots,i_p,I,i_2)}\right)^{L_{i_2}}\times\cdots\times  \left(W_{A^{i_p};(I,i_1,\cdots,i_{p-1})}\right)^{(-1)^pL_{i_p}}=1,
    \label{c2}
\end{equation}
where the indexes $I,(i_1,\cdots,i_p)$ and $[I\neq i_1,\cdots,i_p]$ run cyclically. 
Further, the equations of motions of~\eqref{inbf} ensure that the loop~$W_{a;(i_1,\cdots,i_p)}(\hx_{j_1},\cdots,\hx_{j_{d-p}})$ is deformable so that it does not depend on the spatial coordinate~$\hx_{j_l}$. \par
With the constraints~\eqref{c1} and \eqref{c2}, we count the number of distinct noncontractible loops~\eqref{ap:lpa}. There are~$\binom{d}{p}$
loops which have the form of~$W_{a;(i_1,\cdots,i_p)}$, labeled by $\mathbb{Z}_{\gcd(N,L_{i_1},\cdots,L_{i_p})}$. 
Regarding the loops,~$W_{A^I;(i_1,\cdots,i_p)}$, after some algebra, jointly with the constraints~\eqref{c1} and \eqref{c2}, one finds that there are 
\begin{eqnarray*}
    N^{p\binom{d}{p}}\times\prod_{1\leq i_1<i_2<\cdots<i_{p+1}\leq d}\left[N^p\gcd (N,L_{i_1},L_{i_2},\cdots,L_{i_{p+1}})\right]
\end{eqnarray*}
distinct noncontractible loops.
Therefore, the GSD, which amounts the total number of distinct configurations of loops~\eqref{ap:lpa} is given by
\begin{eqnarray}
   {\rm GSD }
&=& \prod_{1\leq i_1<i_2\cdots <i_p\leq d}\gcd (N,L_{i_1},L_{i_2},\cdots,L_{i_p})\times N^{p\binom{d}{p}}\times\prod_{1\leq i_1<i_2<\cdots<i_{p+1}\leq d}\left[N^p\gcd (N,L_{i_1},L_{i_2},\cdots,L_{i_{p+1}})\right]\nonumber\\
&=&N^{K(d,p)}\times\prod_{1\leq i_1<i_2\cdots <i_p\leq d}\gcd (N,L_{i_1},L_{i_2},\cdots,L_{i_p})\times\prod_{1\leq i_1<i_2\cdots <i_{p+1}\leq d}\gcd (N,L_{i_1},L_{i_2},\cdots,L_{i_{p+1}}),
\end{eqnarray}
which leads to~\eqref{gs} with $K(d,p)$ described
by~\eqref{34}.

\section{Construction of lattice spin models with dipole symmetry}\label{app:st}
In this appendix, we demonstrate several concrete UV spin models that correspond to the dipole BF theory~\eqref{inbf}. We focus on the case of $d=3$ and $p=1,2$. Generalization to other cases of $d$ and $p$ should be straightforward. 
To this end, we follow an approach proposed in~\cite{2024multipole}.
A key insight to construct the spin models is 
to recognize the first and third terms in the BF theory~\eqref{inbf} as the $p$-form BF theories whereas the second term as
as the gauged ``foliated symmetry protected topological~(SPT) phases'', i.e., gauged SPT phases~\cite{spt2013} stacked in layers.
Based on this observation, one can introduce a spin model in a paramagnet phase and implement control phase gate operations on the spins, which corresponds to accommodating the SPT phases. Gauging a global symmetry of the model results in the desired topological spin model associated with the dipole BF theory~\eqref{inbf}. The crucial point is to implement the controlled phase gate operation. Without it, we would end up with the decoupled layers of the spin model corresponding to the~$p$-form BF theories (i.e., decoupled layers of the toric codes~\cite{KITAEV20032}).\par
Recalling the fact that the foliation field $e^I$, which is a one form field along which theories defined on submanifold are stacked,
as well as that the form~$b\fl{d-p}\wedge A\fii p$ is associated with the topological field theoretical description of the SPT phase~\cite{propitius1995topological} with $(d-p)$- and $p$-form global symmetries, 
the coupling term in the BF theory,~$b\fl{d-p}\wedge A\fii p\wedge e^I$ can be interpreted as the stack of gauged~$\mathbb{Z}_N^{(d-p)}\times\mathbb{Z}_N^{(p)}$ SPT phases\footnote{%
The index $\mathbb{Z}_N^{(q)}$ denotes $q$-form $\mathbb{Z}_N$ symmetry.} along the $I$-th direction. 

\subsection{$p=1$ and $d=3$}
To see the construction more explicitly, we envisage a cubic lattice to realize the spin model, which corresponds to $d=3$ and $p=1$. Further, for simplicity, we focus on the case with $N=2$. The generalization to the case of $N>2$ is straightforward by thinking of orientations of the lattice appropriately. On the cubic lattice, we think of qubits each of which resides on links.
Also, we introduce three types of qubits on each vertex of the dual cubic lattice (i.e., three qubits at the center of the cube of the original lattice). Denoting the coordinate of the vertex of the cubic [dual cubic] lattice as $\mathbf{r}\vcentcolon=(\hx,\hy,\hz)$~$\left[\br^*\vcentcolon=(\hx,\hy,\hz)+(\fr,\fr,\fr)\right]$, and vectors to label the links of the lattice as $\bx\vcentcolon=(\fr,0,0)$, $\by\vcentcolon=(0,\fr,0)$, $\bz\vcentcolon=(0,0,\fr)$,
we write the Pauli operators acting on these qubits as $\tau^X_{0,\br+\bi}$, $\tau^Z_{0,\br+\bi}$~($I=x,y,z$), $\tau^X_{i,\br^*}$, $\tau^Z_{i,\br^*}$~($i=1,2,3$), where the first index of the last two operators distinguishes the  three types of the qubits. We dub the qubit (resp:~three qubits) defined on links of the original cubic lattice~(resp:~vertex of the dual cubic lattice) as qubits with type~$0$ (resp:~qubit with type $i=1,2,3$.).\par
With these preparations,
we introduce the following paramagnet Hamiltonian:
\begin{eqnarray}
    H=-\sum_{\br,\br^*}\left[\sum_{I=x,y,z}\tau^X_{0,\br+\bi}+\sum_{i=1,2,3}\tau^X_{i,\br^*}\right].\label{hami}
\end{eqnarray}
We implement the controlled Z~(CZ) gate on the qubits, which amounts to accommodating a stack of SPT phases~\cite{spt2013}.
\begin{figure}[t]
    \begin{center}
      \begin{subfigure}[h]{0.35\textwidth}
  \includegraphics[width=\textwidth]{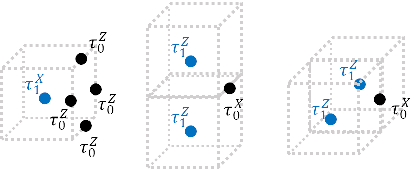}
         \caption{}\label{lt1}
             \end{subfigure}
               \begin{subfigure}[h]{0.35\textwidth}
  \includegraphics[width=\textwidth]{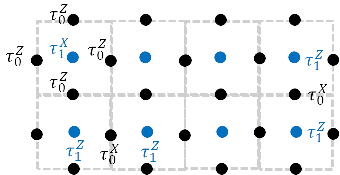}
         \caption{}\label{lt2}
             \end{subfigure}
\end{center}
 \caption{(a) Examples of pairs of qubits on which the CZ gates~\eqref{cz2} act. (b)~The configuration of the qubits viewing from the $x$~axis. The examples of the CZ gates corresponding to left, middle and right of~(a) are also shown
in upper left, right, and bottom left, respectively. 
 }
 \end{figure}
The CZ gate operation which acts on a pair of qubits reads
\begin{equation}
    CZ_{a,b}\ket{a}\ket{b}=(-1)^{ab}\ket{a}\ket{b}\quad(a,b\in\mathbb{Z}_2).
    \label{cz}
\end{equation}
We think of acting the CZ gate on qubits with type~$0$ on the link in the $yz$~plane of the original cubic lattice and the ones with type $1$ at the nodes of the dual cubic lattice which are adjacent to the $yz$-plane. We show an example of such pairs of qubits in Fig.~\eqref{lt1}.
We introduce the following CZ gates:
\begin{eqnarray}
&& \prod_{s,t=\pm1}CZ_{[1,\br^*],[0,\mathbf{f}_{yz}+2\bx+s\by+t\bz]},\nonumber\\ 
&& CZ_{[0,\br+\by],[1,\br^*-2\bx]}CZ_{[0,\br+\by],[1,\br^*-2\bx-2\bz]},\quad CZ_{[0,\br+\bz],[1,\br^*-2\bx]}CZ_{[0,\br+\bz],[1,\br^*-2\bx-2\by]},\label{cz2}
\end{eqnarray}
where the first, second and third terms in~\eqref{cz2} are portrayed in the left, middle, and right of Fig.~\eqref{lt1}.
Here,~$\mathbf{f}_{yz}$ represents the coordinate of a plaquette on $yz$~plane, viz $\mathbf{f}_{yz}\vcentcolon=\br+(0,\fr,\fr)$.
Viewing from the~$x$~axis, the 
CZ gates in~\eqref{cz2} remind us of the cluster states corresponding to the  $\mathbb{Z}_2^{(1)}\times\mathbb{Z}_2^{(0)}$ SPT phase~(see Fig.~\ref{lt2}).
Likewise, we implement CZ gates on qubits with type~2(resp:~3) on $zx(\text{resp:}~xy)$~plane and the ones with type $0$ at the nodes adjacent to the plane. By acting the CZ gates, the Hamiltonian~\eqref{hami} is transformed into 
\begin{eqnarray}
   \widehat{H}
   &=& -\sum_{\br,\br^*}\biggl[\tau^X_{0,\br+\bx}\tau^Z_{2,\br^*-2\by}\tau^Z_{2,\br^*-2\by+2\bz}\tau^Z_{3,\br^*}\tau^Z_{3,\br^*-2\by}
+\tau^X_{0,\br+\by}\tau^Z_{1,\br^*-2\bx}\tau^Z_{1,\br^*-2\bx+2\bz}\tau^Z_{3,\br^*}\tau^Z_{3,\br^*-2\bx}\nonumber\\
&&~~~~~~~~~~  +\tau^X_{0,\br+\bz}\tau^Z_{1,\br^*-2\bx}\tau^Z_{1,\br^*-2\bx+2\by}\tau^Z_{2,\br^*}\tau^Z_{2,\br^*-2\bx}
 +\tau^X_{1,\br^*-2\bx}\prod_{s,t=\pm1}\tau^Z_{0,\mathbf{f}_{yz}+s\by+t\bz} \nonumber\\
&&~~~~~~~~~~ +\tau^X_{1,\br^*-2\by}\prod_{s,t=\pm1}\tau^Z_{0,\mathbf{f}_{zx}+s\bz+t\bx}+\tau^X_{1,\br^*-2\bz}\prod_{s,t=\pm1}\tau^Z_{0,\mathbf{f}_{xy}+s\bx+t\by}\biggr],
\label{hami2}
\end{eqnarray}
where  $\mathbf{f}_{zx}$ and $\mathbf{f}_{xy}$ represent the coordinate of a plaquette on $zx$~and $xy$~plane, respectively, namely ~$\mathbf{f}_{zx}\vcentcolon=\br+(\fr,0,\fr)$, $\mathbf{f}_{xy}\vcentcolon=\br+(\fr,\fr,0)$.
The Hamiltonian~\eqref{hami2} respects $1$-form and $0$-form global symmetries,~$\mathbb{Z}_2^{(1)}\times[\mathbb{Z}_2^{(0)}]^3$, i.e., it is invariant under the following spin flips:
\begin{eqnarray}
\prod_{l\in S^*}\tau^X_{0,l},\quad \prod_{i=1,2,3}\prod_{\br}\tau^X_{i,\br}.\label{sym}
\end{eqnarray}
Here, the first term consists of product of $\tau^X_0$ on links that form a closed dual surface, which we abbreviate as $\prod_{l\in S^*}$.\footnote{Example of a such term is given by $\prod_{s,t,u=\pm1}\tau^X_{0,\br+s\bx+t\by+u\bz}$.}
\par
\begin{figure}[t]
    \begin{center}
         \includegraphics[width=0.65\textwidth]{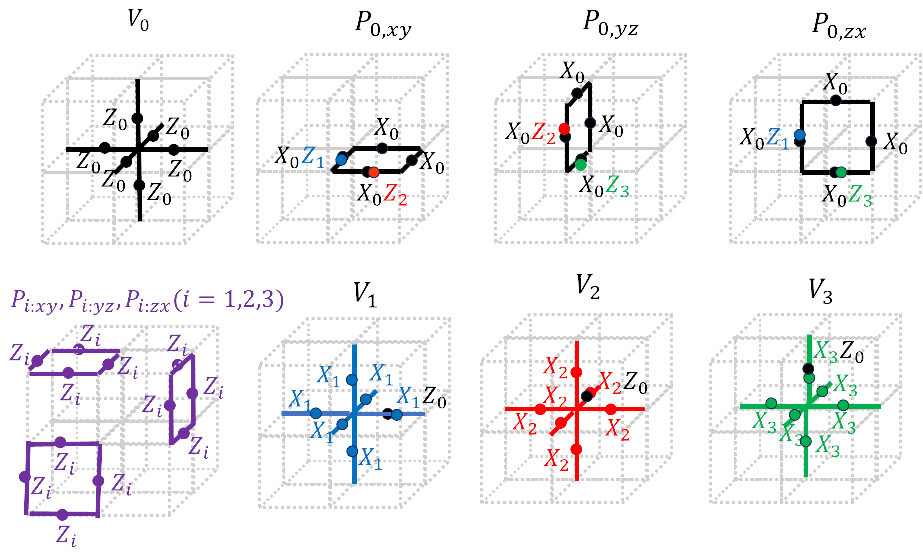}
       \end{center}
       \caption{The spin model which corresponds to~\eqref{inbf} with $N=2$, $d=3$ and $p=1$, described by the Hamiltonian~\eqref{hami3}-\eqref{tcii}. The terms given in~\eqref{tc0} are depicted in the first line whereas the ones in~\eqref{tcii} are in the second line. }
        \label{dip11}
   \end{figure}

Now we are in a good stage to gauge the global symmetries~\eqref{sym}. 
To do so, we promote the global symmetries to local ones by introducing an extended Hilbert space and impose the Gauss law~\cite{Shavit_RevModPhys.52.453,PhysRevB.86.115109}. 
We define states of qubits on each face of the cubic lattice and the ones on the links of the dual cubic lattice, whose Pauli operators are denoted by $X_{0,\mathbf{f}_{ab}}$, $Z_{0,\mathbf{f}_{ab}}$~($ab=xy,yz,zx$), $X_{i,\br^*+\bi}$, $Z_{i,\bi}$. 
Such states are interpreted as the gauge fields. We introduce the following Gauss law operators:
\begin{eqnarray}
    G_{0,\br+\bx}\vcentcolon=\tau^X_{0,\br+\bi}\prod_{\mathbf{f}\supset \bi}X_{0,\mathbf{f}},\quad 
    G_{i,\br^*}\vcentcolon=\tau^X_{i,\br^*}\prod_{\mathbf{l}^*\supset \br^*}X_{i,\mathbf{l}^*}\quad(i=1,2,3).\label{gauss}
\end{eqnarray}
Here, the product $\prod_{\mathbf{f}\supset \bi}$~stands for multiplication of the Pauli operators on plaquettes connected with a link $\br+\bi$. Likewise, we mean the product~$\prod_{\mathbf{l}^*\supset \br^*}$ by multiplication of the Pauli operators on dual links which are connected with a vertex on the dual lattice, $\br^*$.
Using the operators~\eqref{gauss}, we impose that the physical state satisfies 
\begin{eqnarray}
G_{0,\br+\bx}\ket{\text{phys}}=G_{i,\br^*}\ket{\text{phys}}=\ket{\text{phys}}.\label{gauss2}
\end{eqnarray}
Further,
we minimally couple the quartet and quadratic spin coupling terms in the Hamiltonian~\eqref{hami2} to the gauge fields via
\begin{eqnarray}
    \prod_{l\subset \mathbf{f}_{ab}}\tau^Z_{0,l}\to Z_{0,\mathbf{f}_{ab}}\prod_{l\subset \mathbf{f}_{ab}}\tau^Z_{0,l},\quad \tau^Z_{i,\br^*}\tau^Z_{i,\br^*-2\bi}\to \tau^Z_{i,\br^*}Z_{i\br^*-\bi}\tau^Z_{i,\br^*-2\bi}
\end{eqnarray}
so that these terms commute with the Gauss law~\eqref{gauss}. To ensure the fluxless condition making the gauge theory dynamically trivial, we add the following term to the Hamiltonian~\eqref{hami2}:
\begin{eqnarray}
    -\sum_{c}\;\prod_{\mathbf{f}_{ab}\subset c}Z_{0,\mathbf{f}_{ab}}-\sum_{i=1,2,3}\sum_{P^*}\prod_{l^*\subset P^*}Z_{i,l^*},
\end{eqnarray}
where the product in the first~(second) term represents multiplication of the Pauli operators of faces (dual links) that surround a cube (dual plaquette). By the Gauss law~\eqref{gauss2}, we set $\tau^X_{0,\br}=\prod_{\mathbf{f}\supset \bi}X_{0,\mathbf{f}},\tau^X_{i,\br^*}=\prod_{\mathbf{l}^*\supset \br^*}X_{i,\mathbf{l}^*}$. Also, for the sake of making the figure more visually friendly, 
we rearrange the lattice grid so that Pauli operators of the gauge fields are located on links of the newly defined lattice. Overall, we arrive at the following gauged Hamiltonian:
\begin{eqnarray}
    \widehat{H}=-\sum_{\mathbf{f}_{ab},\br}\left(P_{0,\mathbf{f}_{xy}}+P_{0,\mathbf{f}_{yz}}+P_{0,\mathbf{f}_{zx}}+V_{0,\br}\right)-\sum_{i=1,2,3}\sum_{\mathbf{f}_{ab},\br}\left(P_{i,\mathbf{f}_{xy}}+P_{i,\mathbf{f}_{yz}}+P_{i,\mathbf{f}_{zx}}+V_{i,\br}\right)\label{hami3}
\end{eqnarray}
with 
\begin{eqnarray}
&& P_{0,\mathbf{f}_{xy}}\vcentcolon=X_{1,\mathbf{f}_{xy}-\bx}X_{2,\mathbf{f}_{xy}-\by}\prod_{s,t=\pm1}X_{0,\mathbf{f}_{yz}+s\bx+t\by},\quad P_{0,\mathbf{f}_{yz}}\vcentcolon=X_{2,\mathbf{f}_{yz}-\by}X_{3,\mathbf{f}_{yz}-\bz}\prod_{s,t=\pm1}X_{0,\mathbf{f}_{yz}+s\by+t\bz}\nonumber\\
&& P_{0,\mathbf{f}_{zx}}\vcentcolon=X_{3,\mathbf{f}_{zx}-\bz}X_{1,\mathbf{f}_{zx}-\bx}\prod_{s,t=\pm1}X_{0,\mathbf{f}_{zx}+s\bz+t\bx},\quad V_{0,\br}\vcentcolon=\prod_{s,t,u=\pm1}Z_{0,\br+s\bx+t\by+u\bz}\label{tc0}
\end{eqnarray}
and\begin{eqnarray}
&& V_{1,\br}\vcentcolon=Z_{0,\br+\bx}\prod_{s,t,u=\pm1}X_{1,\br+s\bx+t\by+u\bz},\quad V_{2,\br}\vcentcolon=Z_{0,\br+\by}\prod_{s,t,u=\pm1}X_{2,\br+s\bx+t\by+u\bz}\nonumber\\
&& V_{3,\br}\vcentcolon=Z_{0,\br+\bz}\prod_{s,t,u=\pm1}X_{3,\br+s\bx+t\by+u\bz},\quad P_{i,\mathbf{f}_{ab}}\vcentcolon=\prod_{s,t=\pm1}Z_{i,\mathbf{f}_{ab}+s\mathbf{l}_a+t\mathbf{l}_b}~(a\neq b=\{x,y,z\})\label{tcii}
\end{eqnarray}
We portray the terms~\eqref{tc0} and \eqref{tcii} in Fig.~\ref{dip11}. The Hamiltonian~\eqref{hami3} has the similar form as four copies of the $(3+1)$d $\mathbb{Z}_2$ toric codes~\cite{KITAEV20032} with crucial difference being that a few Pauli operators are multiplies with the terms as indicated in the first three terms in~\eqref{tc0} and~\eqref{tcii}, which correspond to the foliated SPT phases. 

The Hamiltonian~\eqref{hami3} is exactly solvable as the individual terms
commute with each other. 
Hence, the ground state is a state with all of the eigenvalues of the terms~\eqref{tc0} and \eqref{tcii} being one.
The role of the gauged foliated SPT phases is to change the statistics of anyons, giving rise to a topological model with dipole symmetry~\cite{2024multipole}.

To see the model~\eqref{hami3} indeed corresponds to the BF theory~\eqref{inbf} with $p=1$ and~$d=3$, we evaluate the GSD on the torus geometry.
 Similar to the case with the toric code~\cite{KITAEV20032}, we count the noncontractible loops or membranes (i.e., product Pauli operators, forming a plane) consisting of the Pauli  operators $Z_0$, $Z_i$,
 that commute with the Hamiltonian~\eqref{hami3} (namely, logical operators) on the $3D$ torus with system size~$L_x\times L_y\times L_z$. 
 One finds the following membrane operators consisting of $Z_0$:
 \begin{eqnarray}
V_{0:xy}\left(\hz+\fr\right)=\prod_{\hx=1}^{L_x}\prod_{\hy=1}^{L_y}Z_{0,\br+\bz},\quad
     V_{0:yz}\left(\hx+\fr\right)=\prod_{\hy=1}^{L_y}\prod_{\hz=1}^{L_z}Z_{0,\br+\bx},\quad 
     V_{0:zx}\left(\hy+\fr\right)=\prod_{\hz=1}^{L_z}\prod_{\hx=1}^{L_x}Z_{0,\br+\by}
     \label{mem}
 \end{eqnarray}
By multiplying the term $V_{0,\br}$, it can be shown that these loops are deformable, implying that they do not depend on the spatial coordinate. The form of these membrane operators are also found in the $(3+1)$d~toric code, yet in the present case there are a few constraints due to the foliated SPT phases.
Indeed, multiplying $V_{1,\br}$ over the entire lattice gives
\begin{eqnarray}
\prod_{\hx=1}^{L_x} V_{0:yz}\left(\hx+\fr\right)=\left( V_{0:yz}\right)^{L_x}=1,\label{m1}
\end{eqnarray}
where we have used the fact that the membrane operators~\eqref{mem} do not depend on the spatial 
coordinate. From~\eqref{m1}, it follows that 
depending on whether~$L_x$ is even or odd, the membrane $W_{0;yz}$ becomes trivial or nontrivial, which can be succinctly described by that 
the membrane operator $W_{0;yz}$ is characterized by $\mathbb{Z}_{\gcd(2,L_x)}$.Similar reasoning can be applied to other membranes in~\eqref{mem}, indicating that the membranes $W_{0:xy}$ and $W_{0:zx}$ is labeled by $\mathbb{Z}_{\gcd(2,L_z)}$, and $\mathbb{Z}_{\gcd(2,L_y)}$, respectively.\par
We turn to evaluation of the loops consisting of string of the Pauli operators $Z_i$. One finds that the model admits the following loop operators:
\begin{eqnarray}
    W_{i;x}(\hy,\hz)=\prod_{\hx=1}^{L_x}Z_{i,\br+\bx},\quad   W_{i;y}(\hz,\hx)=\prod_{\hy=1}^{L_y}Z_{i,\br+\by},\quad   W_{i;z}(\hx,\hy)=\prod_{\hz=1}^{L_z}Z_{i,\br+\bz}\quad(i=1,2,3)\label{kin}
\end{eqnarray}
One can verify that these loops are deformable so that they do not depend on the spatial coordinate.
Naively, there are $2^9$ distinct loops. However, it is incorrect as
we have to take care of several constraints on the loops, stemming from the foliated SPT phases. 
One constraint comes from the fact that multiplying~$P_{0:\mathbf{f}_{xy}}$ on a $xy$ plane gives
\begin{eqnarray}
   \prod_{\hx=1}^{L_x} \prod_{\hy=1}^{L_y}P_{0:\mathbf{f}_{xy}}=1
   \quad \Leftrightarrow \quad   
   W_{1:y}^{L_x}\times W_{2:x}^{L_y}=1.\label{con1}
\end{eqnarray}
Another constraint stems from deformation of other types of loops which consists of string of the Pauli $X_0$ operators dressed with $Z_i$. Indeed, one finds that the model admits the following loop operator:
\begin{eqnarray}
    W_{0:x}(\hy,\hz)=\left[\prod_{\hx=1}^{L_x}X_{0,\br+\bx}\times(Z_{1,\br+\bx})^{\hx}\right]^{\alpha_x},\label{lp}
\end{eqnarray}
where $\alpha_x=\frac{2}{\gcd(2,L_x)}$.
Such a loop was discussed in different spin models in~\cite{2024multipole}.
Deforming the loop~\eqref{lp} in the $y$~direction results in
\begin{eqnarray}
    W_{0:x}(\hy+1,\hz)=W_{0:x}(\hy,\hz)W_{2:x}^{\alpha_x}\label{lpc}
\end{eqnarray}
Iterative use of~\eqref{lpc} and taking the periodic boundary condition into account, we have
\begin{eqnarray}
    W_{2:x}^{\alpha_xL_y}=1.\label{con2}
\end{eqnarray}
By the same token, one can show that
\begin{eqnarray}
     W_{1:y}^{\alpha_yL_x}=1,\label{con3}
\end{eqnarray}
where $\alpha_y=\frac{2}{\gcd(2,L_y)}$.
From the conditions~\eqref{con1}, \eqref{con2} and \eqref{con3}, one finds that there are $2\times \gcd(2,L_x,L_y)$  
distinct configurations of the loops, $W_{1,y}$ and $W_{2,x}$. Likewise, one can show that there are  $2\times \gcd(2,L_y,L_z)$ [$2\times \gcd(2,L_z,L_x)$]  
distinct configurations of the loops, $W_{3,y}$ and $W_{2,z}$ [$W_{3,x}$ and $W_{1,z}$]. There is no constraint on the three loops, $W_{1,x}$, $W_{2:y}$, and $W_{3:z}$, each of which is labeled by $\mathbb{Z}_2$.
Overall, the number of distinct loops~\eqref{kin} is found to be
\begin{eqnarray*}
    2^6\times \gcd(2,L_x,L_y)\times \gcd(2,L_y,L_z)\times \gcd(2,L_z,L_x).
\end{eqnarray*}
Taking the fact that there are $\gcd(2,L_x)\times \gcd(2,L_y)\times \gcd(2,L_z)$ distinct membrane operators~\eqref{mem} into consideration, we finally arrive at that the GSD has the form of~\eqref{gs} in the main text with $N=2$, $p=1$, and $d=3$.

\subsection{$p=2$ and $d=3$}
Similar to the argument in the previous subsection, we can also construct a UV spin model, corresponding to the BF theory~\eqref{inbf} with $p=2$ and $d=3$. Since the way we construct the model closely parallels the one in the previous subsection, we outline how to do it succinctly. \par
%
 \begin{figure}[t]
    \begin{center}
         \includegraphics[width=0.70\textwidth]{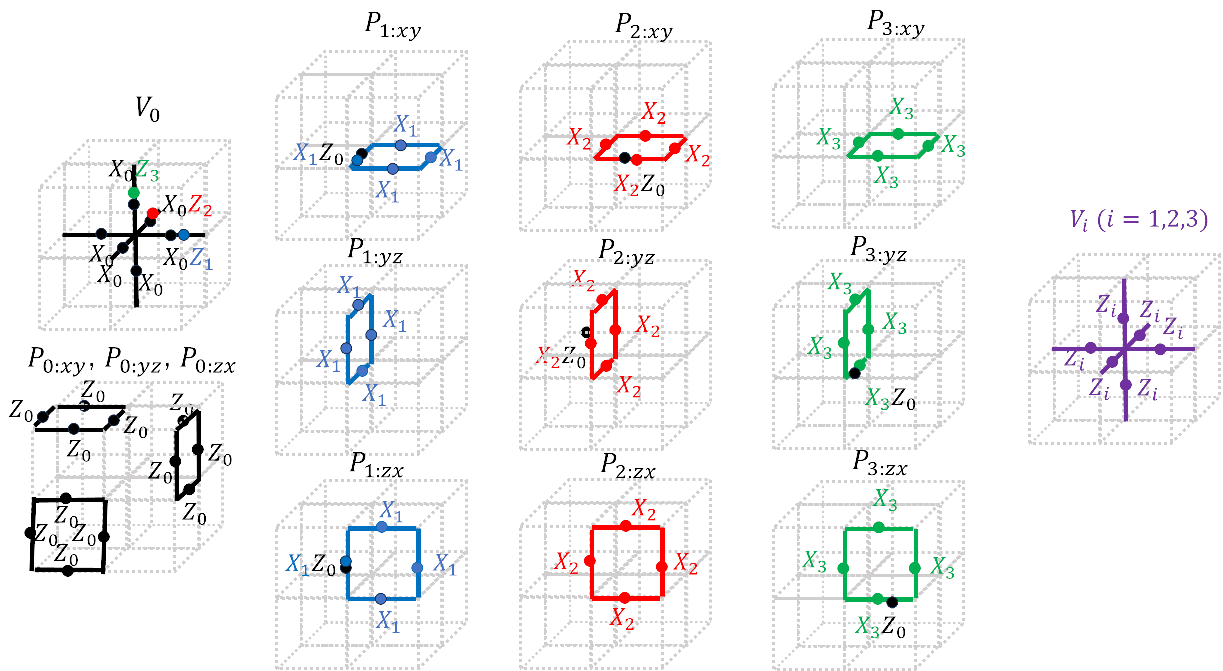}
       \end{center}
       \caption{The spin model which corresponds to the BF theory~\eqref{inbf} with $p=2$, $d=3$, and $N=2$, described by the Hamiltonian~\eqref{hami4}-\eqref{ro3}.
 }
        \label{plane}
   \end{figure}
In the cubic lattice, we introduce qubits on vertices and three types of qubits on the vertices of the dual cubic lattice. Defining the paramagnet Hamiltonian in terms of Pauli $\tau^X$ operators, we implement the CZ gates on it, associated with the foliated SPT phases, respecting global symmetries~$[\mathbb{Z}_2^{(1)}]^3\times \mathbb{Z}_2^{(0)}$. After gauging global symmetries, $[\mathbb{Z}_2^{(1)}]^3\times \mathbb{Z}_2^{(0)}$,
the Hamiltonian is given by (see Fig.~\ref{plane})
\begin{eqnarray}
    \widehat{H}=-\sum_{\mathbf{f}_{ab},\br}\left(P_{0,\mathbf{f}_{xy}}+P_{0,\mathbf{f}_{yz}}+P_{0,\mathbf{f}_{zx}}+V_{0,\br}\right)-\sum_{i=1,2,3}\sum_{\mathbf{f}_{ab},\br}\left(P_{i,\mathbf{f}_{xy}}+P_{i,\mathbf{f}_{yz}}+P_{i,\mathbf{f}_{zx}}+V_{i,\br}\right)\label{hami4}
\end{eqnarray}
with 
\begin{eqnarray}
    V_{0,\br}=Z_{1,\br+\bx}Z_{2,\br+\by}Z_{3,\br+\bz}
\prod_{s,t,u=\pm1}X_{0,\br+s\bx+t\by+u\bz},\quad P_{0,\mathbf{f}_{ab}}\vcentcolon=\prod_{s,t=\pm1}Z_{0,\mathbf{f}_{ab}+s\mathbf{l}_a+t\mathbf{l}_b}~(a\neq b=\{x,y,z\})\label{ro2}
\end{eqnarray}
and 
\begin{eqnarray}
&& P_{1,\mathbf{f}_{xy}}\vcentcolon=Z_{0,\mathbf{f}_{xy}-\bx}\prod_{s,t=\pm1}X_{1,\mathbf{f}_{xy}+s\mathbf{l}_x+t\mathbf{l}_y},\quad P_{1,\mathbf{f}_{yz}}\vcentcolon=\prod_{s,t=\pm1}X_{1,\mathbf{f}_{yz}+s\mathbf{l}_y+t\mathbf{l}_z},\quad P_{1,\mathbf{f}_{zx}}\vcentcolon=Z_{0,\mathbf{f}_{zx}-\bx}\prod_{s,t=\pm1}X_{1,\mathbf{f}_{zx}+s\mathbf{l}_z+t\mathbf{l}_x}\nonumber\\
&& P_{2,\mathbf{f}_{xy}}\vcentcolon=Z_{0,\mathbf{f}_{xy}-\by}\prod_{s,t=\pm1}X_{2,\mathbf{f}_{xy}+s\mathbf{l}_x+t\mathbf{l}_y},\quad P_{2,\mathbf{f}_{yz}}\vcentcolon=Z_{0,\mathbf{f}_{yz}-\by}\prod_{s,t=\pm1}X_{2,\mathbf{f}_{yz}+s\mathbf{l}_y+t\mathbf{l}_z},\quad P_{2,\mathbf{f}_{zx}}\vcentcolon=\prod_{s,t=\pm1}X_{2,\mathbf{f}_{zx}+s\mathbf{l}_z+t\mathbf{l}_x}\nonumber\\
&& P_{3,\mathbf{f}_{xy}}\vcentcolon=\prod_{s,t=\pm1}X_{3,\mathbf{f}_{xy}+s\mathbf{l}_x+t\mathbf{l}_y},\quad P_{3,\mathbf{f}_{yz}}\vcentcolon=Z_{0,\mathbf{f}_{yz}-\bz}\prod_{s,t=\pm1}X_{3,\mathbf{f}_{yz}+s\mathbf{l}_y+t\mathbf{l}_z},\quad P_{3,\mathbf{f}_{zx}}\vcentcolon=Z_{0,\mathbf{f}_{zx}-\bz}\prod_{s,t=\pm1}X_{3,\mathbf{f}_{zx}+s\mathbf{l}_z+t\mathbf{l}_x}\nonumber\\
&& V_{i,\br}\vcentcolon=\prod_{s,t,u=\pm1}X_{i,\br+s\bx+t\by+u\bz}\quad (i=1,2,3).\label{ro3}
\end{eqnarray}
The model has the similar form as the four copies of the toric codes with a crucial difference being that a few Pauli operators are attached with the terms, stemming from the foliated SPT phases. \par
The model~\eqref{hami4} is exactly solvable and by analogous lines of thinking to count the number of distinct noncontractible loops, comprised of $Z_0$ or $Z_i$, which is discussed
in the previous subsection, one finds that the GSD is given by
\begin{eqnarray}
   GSD=   2^8\times \gcd(2,L_x,L_y)\times \gcd(2,L_y,L_z)\times \gcd(2,L_z,L_x)\times  \gcd(2,L_x,L_y,L_z)
\end{eqnarray}
which is consistent with~\eqref{gs} with $p=2$, $d=3$, and $N=2$.

\section{Higher form subsystem BF theory and anomaly inflow}\label{ap3}
In this appendix, we introduce higher form BF theories with subsystem symmetry. Such theories are obtained by imposing an additional constraints on the dipole BF theories~\eqref{pform}. We also discuss anomaly inflow for these theories.
\subsection{Model}
We start by the dipole BF theory~\eqref{pform}. Replacing the gauge field $c\fii{d-p}$ with $dB\fii{d-p-1}\wde e^I$, ($I$ is not summed over), we have
\begin{equation}
   \boxed{ 
\mathcal{L}  =\frac{N}{2\pi}\left[b\fl{d-p}\wde \left(da\fl{p}+(-1)^p\sum_IA\fii{p}\wde e^I\right)+\sum_IB\fii{d-p-1}\wde dA\fii{p}\wde e^I\right]\label{foliated p form bf}}.
\end{equation}
Here, we demand that $1\leq p\leq d-1$. Note that in the case of $p=1$ and $d=3$, the theory~\eqref{foliated p form bf} becomes the foliated BF theory describing the $X$-cube model~\cite{Vijay,foliated1,foliated2}.
Compared with the dipole BF theory~\eqref{inbf}, the theory~\eqref{foliated p form bf} has an additional symmetry -- subsystem symmetry. Indeed, the theory~\eqref{foliated p form bf} is invariant under
\begin{eqnarray}
    A\fii p \ \to\ A\fii p+\gamma\fii{p-1} \wedge e^I\label{diz}
\end{eqnarray}
with $\gamma\fii{p-1}$ being an arbitrary $(p-1)$-form field. 
As shown below, the manifestation of the subsystem symmetry can be seen by noticing that the symmetry~\eqref{diz} puts constraints on the form of gauge invariant extended operators of the gauge fields $a\fl p$ and $A\fii p$ which are analogue of Wilson loops for the $p$-form gauge fields supported on a closed $p$-dimensional spatial submanifold. 
The theory~\eqref{foliated p form bf} respects the following gauge symmetry:
\begin{eqnarray}
&&   a\fl p\to a\fl p+d\Lambda\fl{p-1}+(-1)^{p-1}\sum_I\sigma\fii{p-1} \wde e^I,\quad A\fii p\to A\fii p+d\sigma\fii{p-1}+\gamma\fii{p-1}\wedge e^I\nonumber\\
&&   b\fl{d-p}\to  b\fl{d-p}+d\mu\fl{d-p-1},\quad B\fii{d-p-1}\to B\fii{d-p-1}+d\chi\fii{d-p-2}+(-1)^{p+1}\mu\fl{d-p-1}+g\fl{d-p-2}\wedge e^I. \nonumber\\
\end{eqnarray}
Here, $g\fl{d-p-2}$ denotes an arbitrary $(d-p-2)$-form field. 
Similar to the discussion in~Sec.~\ref{322}, one can evaluate the GSD of~\eqref{foliated p form bf} on a discretized torus geometry with system size $L_{1}\times\cdots\times L_{d}$, 
which is accomplished by counting a distinct number of noncontractible loops of the gauge fields~$a\fl p$ and $A\fii p$ in the spatial direction. 
We do not have noncontractible loop consisting of the gauge field $a\fl p$.\footnote{One naively wonders that similar to the case with the dipole BF theory~\eqref{inbf}, we could construct the loop of $a\fl p$ accompanied by the gauge fields $A\fii p$, such as the one given in~\eqref{a}. However, due to the symmetry~\eqref{diz}, such a loop is not allowed.} As for the loops comprised of $A\fii p$, we have
\begin{eqnarray}
    W_{A^I}(\Sigma_{p})(\hx_I)=\int_{\Sigma_p}A\fii p\quad(\Sigma_p\perp I\text{-th direction}).\label{lpsub}
\end{eqnarray}
Here, $\Sigma_p$ denotes the $p$-form spatial submanifold which
has to be perpendicular to the $I$-th direction due to the symmetry~\eqref{diz}. 
Intuitively,~\eqref{lpsub} can be understood by that on each codimension one layer with coordinate $\hx_I$, which is stacked along the $I$-th direction, noncontractible $p$-form loop is constructed.\footnote{Such a loop is associated with higher form analog of "planon". }
The loop~\eqref{lpsub} is the higher form analog of the Wilson loop defined on each layers of codimension one submanifold, found in the $X$-cube model.\footnote{%
See also~\cite{PhysRevB.101.245134} for construction of the fracton lattice models hosting spatially extended excitations.
}
\par
Naively, there are in total $\binom{d-1}{p}\left(\sum_{i=1}^d L_{i}\right)$ 
such loops, however, there are constraints on these loops. Indeed, equations of motions for the field $b\fl{d-p}_{[0j_1\cdots j_{d-p-1}]}$ gives
\begin{eqnarray}
    \partial_{[i_i}a\fl{p}_{i_2\cdots i_{p+1}]}-(-1)^p\sum_IA\fii{p}_{[i_1\cdots i_p}\delta^I_{i_{p+1}]}=0.
\end{eqnarray}
Performing integration over spatial directions $i_1\cdots i_{p+1}$ of the torus results in
\begin{equation}
  \sum_{x_{i_1}=1}^{L_{i_1}} W_{A^{i_1}}(\Sigma_{[i_2\cdots i_{p+1}]})(\hx_{i_1})- \sum_{x_{i_2}=1}^{L_{i_2}} W_{A^{i_2}}(\Sigma_{[i_1i_3\cdots i_{p+1}]})(\hx_{i_2})+\cdots+(-1)^{p+1} \sum_{x_{i_{p+1}}=1}^{L_{i_{p+1}}} W_{A^{i_{p+1}}}(\Sigma_{[i_1\cdots i_{p}]})(\hx_{i_{p+1}})=0.\label{85}
\end{equation}
Here, $\Sigma_{[i_2\cdots i_{p+1}]}$  in the first term represents spatial submanifold in the $i_2\cdots i_{p+1}$ direction and other terms which have the form~$\Sigma_{[**]}$ are similarly defined.
There are $\binom{d}{p+1}$ constraints on the loops~\eqref{lpsub} described by~\eqref{85}. 
Hence,
the GSD is given by
\begin{equation}
   \boxed{ {\rm GSD}=N^{Q(d,p)} \quad \text{with}\ \ 
   Q(d,p)\vcentcolon= \binom{d-1}{p}\left(\sum_{i=1}^d L_{i}\right)-\binom{d}{p+1}}.
\end{equation}
The theory~\eqref{foliated p form bf} is the higher form analog of the foliated BF theory of the fracton models, such as the~$X$-cube model. Similar to the~$X$-cube model, our theory also exhibits the subextensive GSD.\footnote{As a consistency check, when we set $d=3$ and $p=1$, we have $Q(3,1)=2(L_x+L_y+L_z)-3$, which appears in the GSD of the $X$-cube model~\cite{Vijay} on torus.}
\subsection{Anomaly inflow}
We discuss the anomaly inflow for higher form subsystem symmetry.
Analogous to the previous argument, we gauge the $p$-form symmetries by 
\begin{equation}
\mathcal{L} 
=\frac{N}{2\pi}\biggl[b\fl{d-p}\wde \left(da\fl{p}+(-1)^p\sum_IA\fii{p}\wde e^I-\alpha\fl{p+1}\right)+\sum_IB\fii{d-p-1}\wde\left(dA\fii{p} -\Gamma\fii{p+1}\right)\wde e^I\biggr].
\end{equation}
The theory has the following gauge symmetry:
\begin{eqnarray}
&&      a\fl p\to a\fl p+\lambda_a\fl p,\quad A\fii p \to A\fii p+\lambda_A\fii p\nonumber\\
&&    \alpha\fl{p+1}\to\alpha\fl{p+1}+d\lambda_a\fl p+(-1)^p\lambda_A\fii p\wde e^I,\quad \Gamma\fii{p+1}\to \Gamma\fii{p+1}+d\lambda_A\fii p
\label{gt3}
\end{eqnarray}
with $\lambda_a\fl p$ and $\lambda_A\fii p$ being the $p$-form gauge parameters.\par
Also, one can gauge the $(d-p)$-form symmetries via
\begin{eqnarray}
\mathcal{L}  
=\frac{N}{2\pi}\biggl[b\fl{d-p}\wde \left(da\fl{p}+(-1)^p\sum_IA\fii{p}\wde e^I\right)+\sum_IB\fii{d-p-1}\wde dA\fii{p}\wde e^I\nonumber\\-(-1)^{(d+1)(p+1)}a\fl{p}\wde\beta\fl{d+1-p}-(-1)^{d(p+1)}\sum_IA\fii{p}\wde \Xi\fii{d-p}\wde e^I \biggr].
\end{eqnarray}
This theory has the following gauge symmetry:
\begin{eqnarray}
&&   b\fl{d-p}\to  b\fl{d-p}+\lambda_b\fl{d-p}, \quad B\fii{d-p-1}\to B\fii{d-p-1}+\lambda_B\fii{d-p-1}, \nonumber\\
&&  \beta\fl{d-p+1}\to    \beta\fl{d-p+1}+d\lambda_b\fl{d-p},\quad \Xi\fii{d-p}\to  \Xi\fii{d-p}+d\lambda_B\fii{d-p-1}+(-1)^d\lambda_b\fl{d-p}.
\label{gt4}
\end{eqnarray}
If we try to gauge both of $p$-form and $(d-p)$-form symmetries by thinking of the following theory
\begin{eqnarray}
\mathcal{L} 
&=& \frac{N}{2\pi}\biggl[b\fl{d-p}\wde \left(da\fl{p}+(-1)^p\sum_IA\fii{p}\wde e^I-\alpha\fl{p+1}\right)+\sum_IB\fii{d-p-1}\wde\left(dA\fii{p} -\Gamma\fii{p+1}\right)\wde e^I\nonumber\\
&&   -(-1)^{(d+1)(p+1)}a\fl{p}\wde\beta\fl{d+1-p}-(-1)^{d(p+1)}\sum_IA\fii{p}\wde \Xi\fii{d-p}\wde e^I \biggr],
\label{ppp}
\end{eqnarray}
then we have a problem;~\eqref{ppp} is not invariant under the gauge transformations~\eqref{gt3} and \eqref{gt4}. Indeed, the variant of~$\mathcal{L}$ under the transformations reads
\begin{eqnarray}
\delta \mathcal{L} 
&=&  -\frac{N}{2\pi}\biggl[\lambda_b\fl{d-p}\wde \alpha\fl{p+1} 
+\sum_I\lambda_B\fii{d-p-1}\wde \Gamma^I\wedge e^I+(-1)^{(d+1)(p+1)}\lambda_a\fl{p}\wde \left(\beta\fl{d-p+1}+d\lambda_b\fl{d-p}\right)\nonumber\\
&&~~~~~~   +(-1)^{d(p+1)}\sum_I\lambda_A^I\wde \left( \Xi\fii{d-p}+d\lambda_B\fii{d-p-1}+(-1)^d\lambda_b\fl{d-p}\right)\wde e^I\biggr],
\label{tf}
\end{eqnarray}
signaling the mixed~\tht~anomaly. To cancel the anomaly, we introduced the following terms defined in one dimensional higher:
\begin{eqnarray}
  \mathcal{L}_{\rm inflow} =\mathcal{L}_{0} +\mathcal{L}_{\rm spt}
\label{con}
  \end{eqnarray}
with 
\begin{eqnarray}
  \mathcal{L}_0  
  =\frac{N}{2\pi}\biggl[M\fl{d-p}\wedge\left(d\alpha\fl{p+1}+(-1)^p\sum_I \Gamma\fii{p}\wde e^I\right) +\sum_IN\fii{d-p-1}\wde d\Gamma\fii{p}\wde e^I\nonumber\\
    +O\fl p\wde d\beta\fl{d-p+1}+\sum_IP\fii{p}\wde\left(d\Xi\fii{d-p}+(-1)^{(d-1)}\beta\fl{d-p+1}\right)\wde e^I\biggl],
    \label{kk}
\end{eqnarray}
and
\begin{eqnarray}
   \boxed{
 \mathcal{L}_{\rm spt}  
 =-\frac{N}{2\pi}  \biggl[(-1)^{(p+1)(d+1)}\alpha\fl{p+1}\wedge \beta\fl{d-p+1}+(-1)^{d(p+1)}\sum_I \Gamma\fii{p+1}\wde \Xi\fii{d-p}\wde e^I\biggr]}.
 \label{kk4}
\end{eqnarray}
The bulk terms~\eqref{con} consist of two types of terms: the terms which ensure the flatness conditions on the gauge fields~\eqref{kk} introducing auxiliary fields, $M\fl{d-p}$, $N\fii{d-p-1}$, $O\fl p$, $P\fii p$,  
and the ones that
correspond to invertible phases~\eqref{kk4}. 
Up to total derivative, 
the bulk terms~\eqref{con} respect the gauge symmetries~\eqref{gt3} and \eqref{gt4} jointly with
\begin{eqnarray}
&&   M\fl{d-p}\to  M\fl{d-p}+(-1)^{d-p}\lambda_b\fl{d-p},\quad N\fii{d-p-1}\to N\fii{d-p-1}+(-1)^{d-p-1}\lambda_B\fii{d-p-1}\nonumber\\
&&  O\fl p\to O\fl p+(-1)^{d(p+1)+1}\lambda_a\fl p,\quad  P\fii p\to P\fii p+(-1)^{d(p+1)+d}\lambda_A\fii p.\label{gt5}
\end{eqnarray}
More explicitly, under the transformations~\eqref{gt3}, \eqref{gt4} and \eqref{gt5}, ~\eqref{con} gives
\begin{eqnarray}
\delta\mathcal{L}_{\rm inflow} 
&=& -\frac{N}{2\pi}d\biggl[\lambda_b\fl{d-p}\wde \alpha\fl{p+1} +\sum_I\lambda_B\fii{d-p-1}\wde \Gamma^I\wedge e^I \nonumber\\
&&~~~~~~~~~~~ +(-1)^{(d+1)(p+1)}\lambda_a\fl{p}\wde \left(\beta\fl{d-p+1}+d\lambda_b\fl{d-p}\right)\nonumber\\
&&~~~~~~~~~~~    +(-1)^{d(p+1)}\sum_I\lambda_A^I\wde \left( \Xi\fii{d-p}+d\lambda_B\fii{d-p-1}+(-1)^d\lambda_b\fl{d-p}\right)\wde e^I\biggr].
\label{end}
\end{eqnarray}
The terms inside the large bracket in~\eqref{end} is identical to~\eqref{tf}.
This indicates that the mixed~\tht~anomaly is canceled by~\eqref{con}~with~\eqref{kk} and \eqref{kk4}.
\par
Note that in the previous studies, such as~\cite{Burnell:2021reh,Yamaguchi:2021xeq,Honda:2022shd}, discussed anomaly inflow when gauging subsystem symmetries in tensor gauge theories and exotic field theories. 
Here we discuss the anomaly inflow mechanism for such symmetries, described by foliated field theories. 
We leave elucidation of the relation between the anomaly inflow discussed there
and the present case for future study.

\bibliography{main}

\providecommand{\href}[2]{#2}\begingroup\raggedright\begin{thebibliography}{10}

\bibitem{doi:10.1073/pnas.0803726105}
Z.~Nussinov and G.~Ortiz, ``Sufficient symmetry conditions for topological
  quantum order,'' \href{http://dx.doi.org/10.1073/pnas.0803726105}{{\em
  Proceedings of the National Academy of Sciences} {\bfseries 106} no.~40,
  (2009) 16944--16949}.
  \url{https://www.pnas.org/doi/abs/10.1073/pnas.0803726105}.

\bibitem{nussinov2009symmetry}
Z.~Nussinov and G.~Ortiz, ``{A symmetry principle for topological quantum
  order},'' \href{http://dx.doi.org/10.1016/j.aop.2008.11.002}{{\em Annals
  Phys.} {\bfseries 324} (2009) 977--1057},
  \href{http://arxiv.org/abs/cond-mat/0702377}{{\ttfamily
  arXiv:cond-mat/0702377}}.

\bibitem{gaiotto2015generalized}
D.~Gaiotto, A.~Kapustin, N.~Seiberg, and B.~Willett, ``{Generalized Global
  Symmetries},'' \href{http://dx.doi.org/10.1007/JHEP02(2015)172}{{\em JHEP}
  {\bfseries 02} (2015) 172}, \href{http://arxiv.org/abs/1412.5148}{{\ttfamily
  arXiv:1412.5148 [hep-th]}}.

\bibitem{mcgreevy2023generalized}
J.~McGreevy, ``{Generalized Symmetries in Condensed Matter},''
  \href{http://dx.doi.org/10.1146/annurev-conmatphys-040721-021029}{{\em Ann.
  Rev. Condensed Matter Phys.} {\bfseries 14} (2023) 57--82},
  \href{http://arxiv.org/abs/2204.03045}{{\ttfamily arXiv:2204.03045
  [cond-mat.str-el]}}.

\bibitem{bhardwaj2024lectures}
L.~Bhardwaj, L.~E. Bottini, L.~Fraser-Taliente, L.~Gladden, D.~S.~W. Gould,
  A.~Platschorre, and H.~Tillim, ``{Lectures on generalized symmetries},''
  \href{http://dx.doi.org/10.1016/j.physrep.2023.11.002}{{\em Phys. Rept.}
  {\bfseries 1051} (2024) 1--87},
  \href{http://arxiv.org/abs/2307.07547}{{\ttfamily arXiv:2307.07547
  [hep-th]}}.

\bibitem{chamon}
C.~Chamon, ``Quantum glassiness in strongly correlated clean systems: An
  example of topological overprotection,''
  \href{http://dx.doi.org/10.1103/PhysRevLett.94.040402}{{\em Phys. Rev. Lett.}
  {\bfseries 94} (Jan, 2005) 040402}.
  \url{https://link.aps.org/doi/10.1103/PhysRevLett.94.040402}.

\bibitem{Haah2011}
J.~Haah, ``Local stabilizer codes in three dimensions without string logical
  operators,'' \href{http://dx.doi.org/10.1103/PhysRevA.83.042330}{{\em Phys.
  Rev. A} {\bfseries 83} (Apr, 2011) 042330}.
  \url{https://link.aps.org/doi/10.1103/PhysRevA.83.042330}.

\bibitem{Vijay}
S.~Vijay, J.~Haah, and L.~Fu, ``Fracton topological order, generalized lattice
  gauge theory, and duality,''
  \href{http://dx.doi.org/10.1103/PhysRevB.94.235157}{{\em Phys. Rev. B}
  {\bfseries 94} (Dec, 2016) 235157}.
  \url{https://link.aps.org/doi/10.1103/PhysRevB.94.235157}.

\bibitem{PhysRevX.8.031051}
W.~Shirley, K.~Slagle, Z.~Wang, and X.~Chen, ``Fracton models on general
  three-dimensional manifolds,''
  \href{http://dx.doi.org/10.1103/PhysRevX.8.031051}{{\em Phys. Rev. X}
  {\bfseries 8} (Aug, 2018) 031051}.
  \url{https://link.aps.org/doi/10.1103/PhysRevX.8.031051}.

\bibitem{shirley2019fractional}
W.~Shirley, K.~Slagle, and X.~Chen, ``{Fractional excitations in foliated
  fracton phases},'' \href{http://dx.doi.org/10.1016/j.aop.2019.167922}{{\em
  Annals Phys.} {\bfseries 410} (2019) 167922},
  \href{http://arxiv.org/abs/1806.08625}{{\ttfamily arXiv:1806.08625
  [cond-mat.str-el]}}.

\bibitem{griffin2015scalar}
T.~Griffin, K.~T. Grosvenor, P.~Horava, and Z.~Yan, ``{Scalar Field Theories
  with Polynomial Shift Symmetries},''
  \href{http://dx.doi.org/10.1007/s00220-015-2461-2}{{\em Commun. Math. Phys.}
  {\bfseries 340} no.~3, (2015) 985--1048},
  \href{http://arxiv.org/abs/1412.1046}{{\ttfamily arXiv:1412.1046 [hep-th]}}.

\bibitem{Pretko:2018jbi}
M.~Pretko, ``{The Fracton Gauge Principle},''
  \href{http://dx.doi.org/10.1103/PhysRevB.98.115134}{{\em Phys. Rev. B}
  {\bfseries 98} no.~11, (2018) 115134},
  \href{http://arxiv.org/abs/1807.11479}{{\ttfamily arXiv:1807.11479
  [cond-mat.str-el]}}.

\bibitem{PhysRevX.9.031035}
A.~Gromov, ``Towards classification of fracton phases: The multipole algebra,''
  \href{http://dx.doi.org/10.1103/PhysRevX.9.031035}{{\em Phys. Rev. X}
  {\bfseries 9} (Aug, 2019) 031035}.
  \url{https://link.aps.org/doi/10.1103/PhysRevX.9.031035}.

\bibitem{PhysRevB.98.035111}
H.~Ma, M.~Hermele, and X.~Chen, ``Fracton topological order from the higgs and
  partial-confinement mechanisms of rank-two gauge theory,''
  \href{http://dx.doi.org/10.1103/PhysRevB.98.035111}{{\em Phys. Rev. B}
  {\bfseries 98} (Jul, 2018) 035111}.
  \url{https://link.aps.org/doi/10.1103/PhysRevB.98.035111}.

\bibitem{PhysRevB.106.045112}
P.~Gorantla, H.~T. Lam, N.~Seiberg, and S.-H. Shao, ``Global dipole symmetry,
  compact lifshitz theory, tensor gauge theory, and fractons,''
  \href{http://dx.doi.org/10.1103/PhysRevB.106.045112}{{\em Phys. Rev. B}
  {\bfseries 106} (Jul, 2022) 045112}.
  \url{https://link.aps.org/doi/10.1103/PhysRevB.106.045112}.

\bibitem{Jain:2021ibh}
A.~Jain and K.~Jensen, ``{Fractons in curved space},''
  \href{http://dx.doi.org/10.21468/SciPostPhys.12.4.142}{{\em SciPost Phys.}
  {\bfseries 12} no.~4, (2022) 142},
  \href{http://arxiv.org/abs/2111.03973}{{\ttfamily arXiv:2111.03973
  [hep-th]}}.

\bibitem{hirono2022symmetry}
Y.~Hirono, M.~You, S.~Angus, and G.~Y. Cho, ``{A symmetry principle for gauge
  theories with fractons},''
  \href{http://dx.doi.org/10.21468/SciPostPhys.16.2.050}{{\em SciPost Phys.}
  {\bfseries 16} no.~2, (2024) 050},
  \href{http://arxiv.org/abs/2207.00854}{{\ttfamily arXiv:2207.00854
  [cond-mat.str-el]}}.

\bibitem{PhysRevB.97.235112}
D.~Bulmash and M.~Barkeshli, ``Higgs mechanism in higher-rank symmetric u(1)
  gauge theories,'' \href{http://dx.doi.org/10.1103/PhysRevB.97.235112}{{\em
  Phys. Rev. B} {\bfseries 97} (Jun, 2018) 235112}.
  \url{https://link.aps.org/doi/10.1103/PhysRevB.97.235112}.

\bibitem{ebisu2209anisotropic}
H.~Ebisu and B.~Han, ``{Anisotropic higher rank $\mathbb{Z}_N$ topological
  phases on graphs},''
  \href{http://dx.doi.org/10.21468/SciPostPhys.14.5.106}{{\em SciPost Phys.}
  {\bfseries 14} no.~5, (2023) 106},
  \href{http://arxiv.org/abs/2209.07987}{{\ttfamily arXiv:2209.07987
  [cond-mat.str-el]}}.

\bibitem{delfino2023anyon}
G.~Delfino and Y.~You, ``{Anyon condensation web and multipartite entanglement
  in two-dimensional modulated gauge theories},''
  \href{http://dx.doi.org/10.1103/PhysRevB.109.205146}{{\em Phys. Rev. B}
  {\bfseries 109} no.~20, (2024) 205146},
  \href{http://arxiv.org/abs/2310.09490}{{\ttfamily arXiv:2310.09490
  [cond-mat.str-el]}}.

\bibitem{han2024dipolar}
J.~H. Han, ``Dipolar bf theory and dipolar braiding statistics,''
  \href{http://arxiv.org/abs/2403.08158}{{\ttfamily arXiv:2403.08158
  [cond-mat.str-el]}}.

\bibitem{2023foliated}
H.~Ebisu, M.~Honda, and T.~Nakanishi, ``Foliated field theories and multipole
  symmetries,'' \href{http://dx.doi.org/10.1103/PhysRevB.109.165112}{{\em Phys.
  Rev. B} {\bfseries 109} (Apr, 2024) 165112}.
  \url{https://link.aps.org/doi/10.1103/PhysRevB.109.165112}.

\bibitem{2024multipole}
H.~Ebisu, M.~Honda, and T.~Nakanishi, ``Multipole and fracton topological order
  via gauging foliated symmetry protected topological phases,''
  \href{http://dx.doi.org/10.1103/PhysRevResearch.6.023166}{{\em Phys. Rev.
  Res.} {\bfseries 6} (May, 2024) 023166}.
  \url{https://link.aps.org/doi/10.1103/PhysRevResearch.6.023166}.

\bibitem{foliated1}
K.~Slagle, D.~Aasen, and D.~Williamson, ``{Foliated field theory and
  string-membrane-net condensation picture of fracton order},''
  \href{http://dx.doi.org/10.21468/SciPostPhys.6.4.043}{{\em SciPost Phys.}
  {\bfseries 6} (2019) 043}.
  \url{https://scipost.org/10.21468/SciPostPhys.6.4.043}.

\bibitem{foliated2}
K.~Slagle, ``Foliated quantum field theory of fracton order,''
  \href{http://dx.doi.org/10.1103/PhysRevLett.126.101603}{{\em Phys. Rev.
  Lett.} {\bfseries 126} (Mar, 2021) 101603}.
  \url{https://link.aps.org/doi/10.1103/PhysRevLett.126.101603}.

\bibitem{PhysRevB.66.054526}
A.~Paramekanti, L.~Balents, and M.~P.~A. Fisher, ``Ring exchange, the exciton
  bose liquid, and bosonization in two dimensions,''
  \href{http://dx.doi.org/10.1103/PhysRevB.66.054526}{{\em Phys. Rev. B}
  {\bfseries 66} (Aug, 2002) 054526}.
  \url{https://link.aps.org/doi/10.1103/PhysRevB.66.054526}.

\bibitem{seiberg2021exotic}
N.~Seiberg and S.-H. Shao, ``{Exotic Symmetries, Duality, and Fractons in
  2+1-Dimensional Quantum Field Theory},''
  \href{http://dx.doi.org/10.21468/SciPostPhys.10.2.027}{{\em SciPost Phys.}
  {\bfseries 10} no.~2, (2021) 027},
  \href{http://arxiv.org/abs/2003.10466}{{\ttfamily arXiv:2003.10466
  [cond-mat.str-el]}}.

\bibitem{Ohmori:2022rzz}
K.~Ohmori and S.~Shimamura, ``{Foliated-exotic duality in fractonic BF
  theories},'' \href{http://dx.doi.org/10.21468/SciPostPhys.14.6.164}{{\em
  SciPost Phys.} {\bfseries 14} no.~6, (2023) 164},
  \href{http://arxiv.org/abs/2210.11001}{{\ttfamily arXiv:2210.11001
  [hep-th]}}.

\bibitem{Shimamura:2024kwf}
S.~Shimamura, ``{Anomaly of Subsystem Symmetries in Exotic and Foliated $BF$
  Theories},'' \href{http://arxiv.org/abs/2404.10601}{{\ttfamily
  arXiv:2404.10601 [cond-mat.str-el]}}.

\bibitem{CALLAN1985427}
C.~Callan and J.~Harvey, ``Anomalies and fermion zero modes on strings and
  domain walls,''
  \href{http://dx.doi.org/https://doi.org/10.1016/0550-3213(85)90489-4}{{\em
  Nuclear Physics B} {\bfseries 250} no.~1, (1985) 427--436}.
  \url{https://www.sciencedirect.com/science/article/pii/0550321385904894}.

\bibitem{Cao:2023rrb}
W.~Cao and Q.~Jia, ``{Symmetry TFT for subsystem symmetry},''
  \href{http://dx.doi.org/10.1007/JHEP05(2024)225}{{\em JHEP} {\bfseries 05}
  (2024) 225}, \href{http://arxiv.org/abs/2310.01474}{{\ttfamily
  arXiv:2310.01474 [hep-th]}}.

\bibitem{Burnell:2021reh}
F.~J. Burnell, T.~Devakul, P.~Gorantla, H.~T. Lam, and S.-H. Shao, ``{Anomaly
  inflow for subsystem symmetries},''
  \href{http://dx.doi.org/10.1103/PhysRevB.106.085113}{{\em Phys. Rev. B}
  {\bfseries 106} no.~8, (2022) 085113},
  \href{http://arxiv.org/abs/2110.09529}{{\ttfamily arXiv:2110.09529
  [cond-mat.str-el]}}.

\bibitem{Yamaguchi:2021xeq}
S.~Yamaguchi, ``{Gapless edge modes in (4+1)-dimensional topologically massive
  tensor gauge theory and anomaly inflow for subsystem symmetry},''
  \href{http://dx.doi.org/10.1093/ptep/ptac032}{{\em PTEP} {\bfseries 2022}
  no.~3, (2022) 033B08}, \href{http://arxiv.org/abs/2110.12861}{{\ttfamily
  arXiv:2110.12861 [hep-th]}}.

\bibitem{Honda:2022shd}
M.~Honda and T.~Nakanishi, ``{Scalar, fermionic and supersymmetric field
  theories with subsystem symmetries in d + 1 dimensions},''
  \href{http://dx.doi.org/10.1007/JHEP03(2023)188}{{\em JHEP} {\bfseries 03}
  (2023) 188}, \href{http://arxiv.org/abs/2212.13006}{{\ttfamily
  arXiv:2212.13006 [hep-th]}}.

\bibitem{okuda2024anomaly}
T.~Okuda, A.~Parayil~Mana, and H.~Sukeno, ``{Anomaly inflow for CSS and
  fractonic lattice models and dualities via cluster state measurement},''
  \href{http://arxiv.org/abs/2405.15853}{{\ttfamily arXiv:2405.15853
  [quant-ph]}}.

\bibitem{lam2024topological}
H.~T. Lam, J.~H. Han, and Y.~You, ``{Topological Dipole Insulator},''
  \href{http://arxiv.org/abs/2403.13880}{{\ttfamily arXiv:2403.13880
  [cond-mat.mes-hall]}}.

\bibitem{Gorantla:2021svj}
P.~Gorantla, H.~T. Lam, N.~Seiberg, and S.-H. Shao, ``{A modified Villain
  formulation of fractons and other exotic theories},''
  \href{http://dx.doi.org/10.1063/5.0060808}{{\em J. Math. Phys.} {\bfseries
  62} no.~10, (2021) 102301}, \href{http://arxiv.org/abs/2103.01257}{{\ttfamily
  arXiv:2103.01257 [cond-mat.str-el]}}.

\bibitem{KITAEV20032}
A.~Kitaev, ``Fault-tolerant quantum computation by anyons,''
  \href{http://dx.doi.org/https://doi.org/10.1016/S0003-4916(02)00018-0}{{\em
  Annals of Physics} {\bfseries 303} no.~1, (2003) 2--30}.
  \url{https://www.sciencedirect.com/science/article/pii/S0003491602000180}.

\bibitem{PhysRevB.72.035307}
A.~Hamma, P.~Zanardi, and X.-G. Wen, ``String and membrane condensation on
  three-dimensional lattices,''
  \href{http://dx.doi.org/10.1103/PhysRevB.72.035307}{{\em Phys. Rev. B}
  {\bfseries 72} (Jul, 2005) 035307}.
  \url{https://link.aps.org/doi/10.1103/PhysRevB.72.035307}.

\bibitem{tensor_gauge}
M.~Pretko, ``Generalized electromagnetism of subdimensional particles: A spin
  liquid story,'' \href{http://dx.doi.org/10.1103/PhysRevB.96.035119}{{\em
  Phys. Rev. B} {\bfseries 96} (Jul, 2017) 035119}.
  \url{https://link.aps.org/doi/10.1103/PhysRevB.96.035119}.

\bibitem{Cordova:2018cvg}
C.~C\'ordova, T.~T. Dumitrescu, and K.~Intriligator, ``{Exploring 2-Group
  Global Symmetries},'' \href{http://dx.doi.org/10.1007/JHEP02(2019)184}{{\em
  JHEP} {\bfseries 02} (2019) 184},
  \href{http://arxiv.org/abs/1802.04790}{{\ttfamily arXiv:1802.04790
  [hep-th]}}.

\bibitem{pai2018fractonic}
S.~Pai and M.~Pretko, ``Fractonic line excitations: An inroad from
  three-dimensional elasticity theory,''
  \href{http://dx.doi.org/10.1103/physrevb.97.235102}{{\em Physical Review B}
  {\bfseries 97} no.~23, (June, 2018) 235102}.
  \url{http://dx.doi.org/10.1103/PhysRevB.97.235102}.

\bibitem{shenoy2020k}
V.~B. Shenoy and R.~Moessner, ``(k, n)-fractonic maxwell theory,''
  \href{http://dx.doi.org/10.1103/physrevb.101.085106}{{\em Physical Review B}
  {\bfseries 101} no.~8, (Feb., 2020) 085106}.
  \url{http://dx.doi.org/10.1103/PhysRevB.101.085106}.

\bibitem{rbh_PhysRevA.71.062313}
R.~Raussendorf, S.~Bravyi, and J.~Harrington, ``Long-range quantum entanglement
  in noisy cluster states,''
  \href{http://dx.doi.org/10.1103/PhysRevA.71.062313}{{\em Phys. Rev. A}
  {\bfseries 71} (Jun, 2005) 062313}.
  \url{https://link.aps.org/doi/10.1103/PhysRevA.71.062313}.

\bibitem{dSPT}
H.~T. Lam, ``Classification of dipolar symmetry-protected topological phases:
  Matrix product states, stabilizer hamiltonians, and finite tensor gauge
  theories,'' \href{http://dx.doi.org/10.1103/PhysRevB.109.115142}{{\em Phys.
  Rev. B} {\bfseries 109} (Mar, 2024) 115142}.
  \url{https://link.aps.org/doi/10.1103/PhysRevB.109.115142}.

\bibitem{PhysRevB.107.125154}
H.~Ebisu, ``Symmetric higher rank topological phases on generic graphs,''
  \href{http://dx.doi.org/10.1103/PhysRevB.107.125154}{{\em Phys. Rev. B}
  {\bfseries 107} (Mar, 2023) 125154}.
  \url{https://link.aps.org/doi/10.1103/PhysRevB.107.125154}.

\bibitem{spt2013}
X.~Chen, Z.-C. Gu, Z.-X. Liu, and X.-G. Wen, ``Symmetry protected topological
  orders and the group cohomology of their symmetry group,''
  \href{http://dx.doi.org/10.1103/PhysRevB.87.155114}{{\em Phys. Rev. B}
  {\bfseries 87} (Apr, 2013) 155114}.
  \url{https://link.aps.org/doi/10.1103/PhysRevB.87.155114}.

\bibitem{propitius1995topological}
M.~D.~F. de~Wild~Propitius, {\em {Topological interactions in broken gauge
  theories}}.
\newblock PhD thesis, Amsterdam U., 1995.
\newblock \href{http://arxiv.org/abs/hep-th/9511195}{{\ttfamily
  arXiv:hep-th/9511195}}.

\bibitem{Shavit_RevModPhys.52.453}
R.~Savit, ``Duality in field theory and statistical systems,''
  \href{http://dx.doi.org/10.1103/RevModPhys.52.453}{{\em Rev. Mod. Phys.}
  {\bfseries 52} (Apr, 1980) 453--487}.
  \url{https://link.aps.org/doi/10.1103/RevModPhys.52.453}.

\bibitem{PhysRevB.86.115109}
M.~Levin and Z.-C. Gu, ``Braiding statistics approach to symmetry-protected
  topological phases,''
  \href{http://dx.doi.org/10.1103/PhysRevB.86.115109}{{\em Phys. Rev. B}
  {\bfseries 86} (Sep, 2012) 115109}.
  \url{https://link.aps.org/doi/10.1103/PhysRevB.86.115109}.

\bibitem{PhysRevB.101.245134}
M.-Y. Li and P.~Ye, ``Fracton physics of spatially extended excitations,''
  \href{http://dx.doi.org/10.1103/PhysRevB.101.245134}{{\em Phys. Rev. B}
  {\bfseries 101} (Jun, 2020) 245134}.
  \url{https://link.aps.org/doi/10.1103/PhysRevB.101.245134}.

\end{thebibliography}\endgroup
\bibliographystyle{utphys}

\end{document}